\documentclass[letterpaper,twocolumn,10pt]{article}
\usepackage{zhanggroup}

\usepackage{xcolor}
\usepackage[dvipsnames]{xcolor}
\newcommand{\mypara}[1]{\noindent{\bf {#1}.} \xspace}
\newcommand{\OurMethod}{OtomeSCAN\xspace}
\usepackage[utf8]{inputenc}    
\usepackage{CJKutf8}
\makeatletter
\input{c70gbsn.fd}
\DeclareFontShape{C70}{gbsn}{b}{n}
  {<-> CJKb * gbsnu}{\CJKbold}
\makeatother        
\usepackage{tabularx}          
\usepackage{booktabs}          
\usepackage{makecell}
\usepackage{multirow} 
\usepackage[table]{xcolor}
\usepackage{colortbl}
\usepackage{pifont}
\usepackage{xspace}  
\usepackage{amsmath}
\usepackage{subcaption}
\usepackage{siunitx}
\usepackage{url}
\usepackage{hyperref}
\newcommand{\AnnotatedDatasetURL}{\url{https://huggingface.co/datasets/TrustAIRLab/OtomeSCAN}}
\usepackage{tablefootnote}
\definecolor{first}{RGB}{153, 204, 153}
\definecolor{second}{RGB}{210, 240, 210}
\definecolor{third}{RGB}{255, 242, 204} 
\definecolor{gray}{RGB}{230, 230, 230} 
\definecolor{red}{RGB}{227, 23, 13} 
\definecolor{gao}{RGB} {255,220,220}
\definecolor{di}{RGB} {220,255,220}   
\definecolor{zhong}{RGB}{255,255,200} 
\definecolor{negcoef}{RGB}{245,215,220}
\definecolor{poscoef}{RGB}{210,240,210}
\usepackage[T1]{fontenc}
\usepackage{fvextra}
\usepackage[most]{tcolorbox}

\newcommand{\emoji}[1]{\csname emoji#1\endcsname\xspace}
\usepackage{makecell} 
\usepackage{booktabs}
\usepackage{threeparttable}
\usepackage{enumitem}
\usepackage{graphicx}
\usepackage{threeparttable}
\usepackage[utf8]{inputenc}
\usepackage{forest}
\usepackage{CJKutf8}
\usepackage{tikz}
\usetikzlibrary{arrows.meta}

\newcommand{\refappendix}[1]{\hyperref[#1]{Appendix~\ref*{#1}}}

\begin{document}

\date{}

\title{\bf ``Shut Up and Let Me Enjoy My Otome'': Understanding and Measuring the Toxicity in Otome Game Communities}

\author{
Yage Zhang\textsuperscript{1}\ \ \
Xinyue Shen\textsuperscript{1,2}\ \ \
Yukun Jiang\textsuperscript{1}\ \ \
Michael Backes\textsuperscript{1}\ \ \
Yang Zhang\textsuperscript{1}\thanks{Corresponding author.}
\\
\\
\textsuperscript{1}\textit{CISPA Helmholtz Center for Information Security} \ \ \
\textsuperscript{2}\textit{University of Waterloo}
}

\maketitle

\begin{abstract}
Otome games, a romance simulation genre primarily targeting female, have emerged as a major force in the global gaming market, attracting hundreds of millions of players and billions in revenue. 
Despite their popularity, otome game communities face pervasive online toxicity, which has been largely unexplored.
In this work, we present the first large-scale measurement of toxicity in otome game communities across social platforms. 
We introduce \OurMethod, a framework for collecting, evaluating, and analyzing 620,045 posts from Weibo and Reddit spanning 18 months. 
To support robust analysis, we manually annotated a ground-truth dataset of 4,308 posts, identifying eight target groups such as players and game developers.
We evaluate seven toxicity detectors on the dataset, including general-purpose models and our proposed LLM-based detectors, with our best model achieving F1-scores of 0.82 (Weibo) and 0.78 (Reddit). 
Our analysis reveals significant platform-based differences in toxicity:
22.20\% of otome-related posts on Weibo are toxic, compared to 3.71\% on Reddit.
Besides, real-world events like in-community conflicts can rapidly escalate toxicity, with toxicity ratios increasing to 37.09\% in just 72 hours during an external attack on Weibo.
We also flag 191 potential-coordination clusters in otome game communities, 64.40\% of which target game developers, with several accounts participating repeatedly across multiple clusters.
We hope our work inspires further research on community-specific toxicity and contributes to building healthier online spaces for marginalized gaming communities.\footnote{Our dataset is available at \AnnotatedDatasetURL.}

\noindent\textbf{\textcolor{red}{Disclaimer: This paper contains examples of toxic and abusive language.
Reader discretion is recommended.}}
\end{abstract}

\section{Introduction}

\begin{figure}[t!]
    \centering
    \includegraphics[width=1\linewidth]{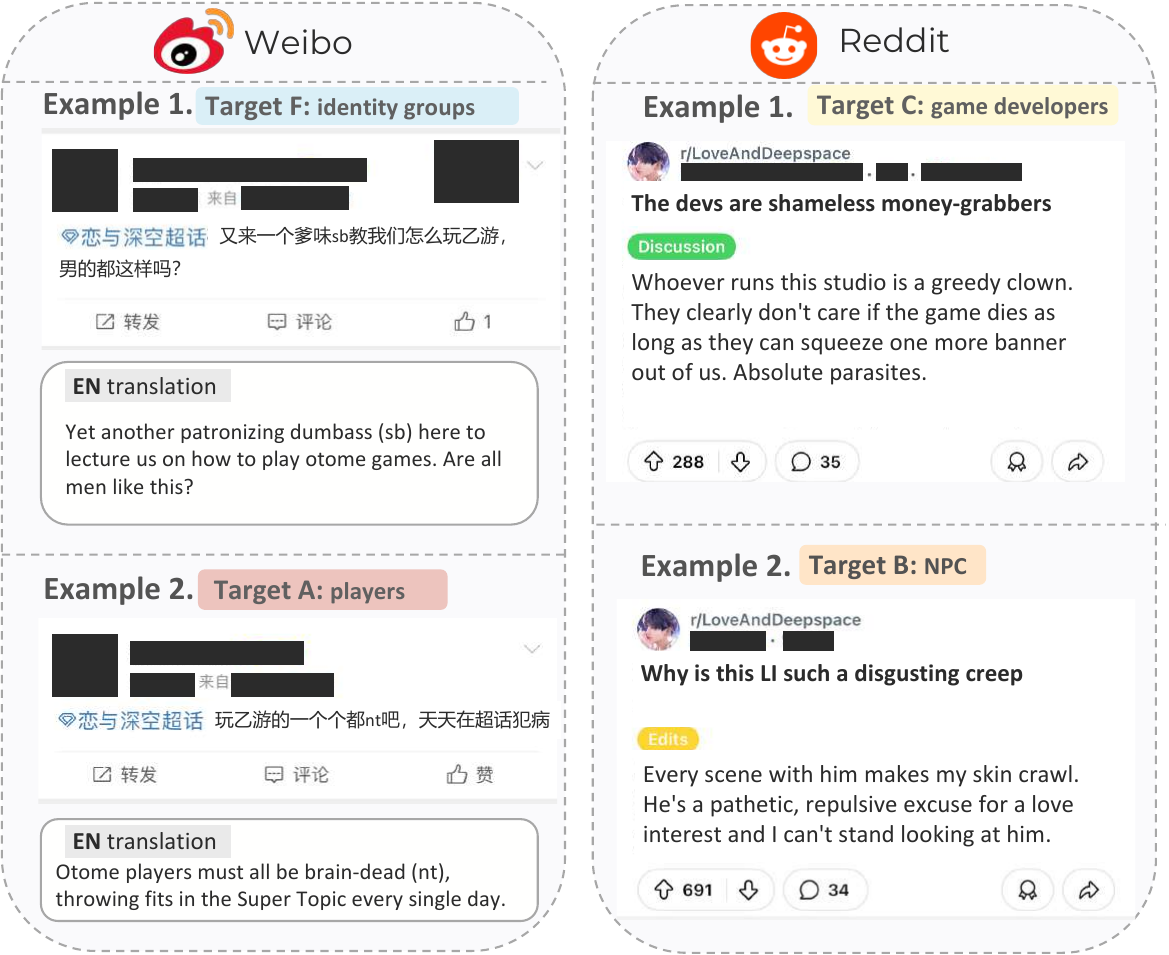}
    \caption{Examples of toxic posts and target groups for otome games on Weibo (left) and Reddit (right).
    We paraphrased the examples to prevent verbatim searches from identifying the users while preserving the original meaning.
    }
    \label{figure:examples}
\end{figure}

Digital games have become an important component of everyday leisure for hundreds of millions of people worldwide.
Within this landscape, online toxicity in gaming communities has attracted growing research attention, with studies documenting gender-based hostility~\cite{TABBBCDDKKMMRS21}, coordinated harassment~\cite{PHTTFM21}, and hate speech dynamics in multiplayer environments~\cite{KK15, K20}.
However, existing work overwhelmingly focuses on general mixed-gender games, leaving ecosystems with fundamentally different social structures largely unexplored.

\textit{Otome games}, a narrative-driven romance simulation genre in which players take on the role of a female protagonist and develop romantic relationships with non-player characters (NPCs)~\cite{PCFLX25}, represent one such ecosystem.
According to recent reports, the global otome games market reached approximately USD~5.26 billion in 2024~\cite{OtomeMarketReport2025}.
One flagship title, \textit{Love and Deepspace}, reportedly reached 50 million global users by early 2025~\cite{LoveAndDeepspaceDating}.
Yet, otome games continue to face widespread online toxicity and discrimination, as illustrated in~\autoref{figure:examples}.
In mainstream gaming communities, otome games are frequently criticized for encouraging women to challenge traditional gender norms~\cite{G24}.
Within otome game communities, in-group harassment and interpersonal hostility are common~\cite{OtomeElitism, OtomeCommunityRidicule}.
A notable incident occurred in August 2024~\cite{LoveAndDeepspaceCourtCase, OtomeFandomManagement}, when a rapper released a satirical song mocking otome game players on the social platform. 
The song quickly went viral, reaching millions of users, and was followed by a surge of toxic posts towards otome game players, rising from 22.20\% to 37.09\% within 72 hours (see~\autoref{section:temporal}).
However, the research community still lacks a systematic understanding of toxicity in otome game communities, including its prevalence, targeted groups, temporal dynamics, and linguistic characteristics.
This gap significantly hinders efforts to address and mitigate online toxicity faced by the otome game players, primarily millions of female players.

\mypara{Our Work}
In this work, we present the first large-scale measurement study on toxicity in otome game communities.
Specifically, we focus on the following research questions:
\begin{itemize}
    \item\textbf{RQ1:} How does toxicity in otome game communities differ from patterns observed in general game communities, in terms of prevalence, target groups, and interaction patterns, and how do the themes of toxic disputes differ across platforms?
    \item\textbf{RQ2:} What types of events are associated with significant peaks of toxicity in otome game communities?
    \item\textbf{RQ3:} What linguistic features are present in toxic posts from otome game communities, and how do they evade platform moderation?
    \item\textbf{RQ4:} Beyond individual toxic posts, what patterns of potential coordination appear in otome game communities? Who are their primary targets?
\end{itemize}

To answer these questions, we introduce \OurMethod, a framework designed for collecting, evaluating, and analyzing the toxicity
in otome game communities.
Leveraging \OurMethod, we collect 620,045 posts from Weibo and Reddit, covering four otome game communities and two general game communities (used later as control groups), spanning from January 2024 to May 2025. 
Given the lack of prior work evaluating the performance of toxicity detectors on otome game content, we randomly sampled and manually annotated 4,308 posts to serve as a ground truth dataset.
This annotation includes two levels of labels: (1) binary toxicity (toxic or non-toxic) and (2) target groups for toxic posts.
In the end, we identified eight target groups in the otome-related toxic posts, which are players, NPCs, game developers, platform moderators, policymakers, identity groups, other game-related entities, and unknown (see~\autoref{section:annotation_process}).
We then evaluate three general-purpose toxicity detectors, i.e., Perspective API, OpenAI Moderation API, COLD, and four of our proposed LLM-driven detectors on the annotated set.
Our best-performing model achieved F1-scores of 0.82 on Weibo and 0.78 on Reddit, significantly outperforming the general-purpose detectors.
We then employ it to annotate the full dataset (see~\autoref{section:detector_evaluation}).

Regarding analysis, we start by performing a comparative analysis of toxicity in otome game communities and general game communities, focusing on the prevalence, target groups, and user interaction patterns across social platforms, and comparing the themes of toxic disputes between the two platforms (RQ1).
We then conduct a time series analysis of toxic posts to identify real-world events that coincide with significant toxicity peaks (RQ2).
Through linguistic analysis, we investigate toxic spans in toxic posts and the variation strategies they contain, which may help toxic content evade platform moderation (RQ3).
Finally, we apply a similarity-based detection procedure to flag candidate clusters exhibiting potential coordination in otome game communities (RQ4).

\mypara{Main Findings}
We make the following main findings:
\begin{itemize}
\item Compared with general game communities, otome game communities show platform-specific differences in both prevalence and target distribution. 
On Weibo, otome discussions are far more toxic than general gaming discussions (22.20\% vs. 3.43\%), with a distinctively higher toxicity ratio targeting players (47.2\% vs. 27.4\%). 
On Reddit, overall toxicity ratios are similar (3.71\% vs. 3.67\%), but otome toxicity shifts toward NPCs rather than game companies (see \autoref{section:RQ1}).
    \item Real-world events like in-community conflicts, game updates, external attacks, and consumer rights protests are associated with significant toxicity surges in otome game communities.
    Across several events, the distribution of targeted groups shifts over time and increasingly includes players (see~\autoref{section:temporal}).
    \item Toxic posts in otome game communities show distinct linguistic patterns across platforms. 
    39.89\% of model-detected unique toxic spans on Weibo involve variation strategies like slang, abbreviation, and substitution, compared with 22.77\% on Reddit, where toxicity is expressed more directly (see~\autoref{section:Linguistic-Strategies}).
    \item Our procedure flags 191 potential-coordination clusters in otome game communities, of which 64.40\% target game developers.
Several accounts participate repeatedly across clusters (see~\autoref{section:case_study_weibo}).
\end{itemize}

\begin{CJK}{UTF8}{gbsn}
\begin{table*}[t!]
\centering
\caption{
Commercial scale (USD) and social‐media engagement of popular otome games~\cite{SensorTower}. 
We collect Weibo and Reddit post counts on June 7, 2025, and they represent cumulative engagement up to that date.
}
\label{table:games_overview}
\scalebox{0.8}{
\begin{tabular}{@{}l l c c c@{}}
\toprule
\textbf{Game Title (English)}  & \textbf{Game Title (Chinese)} & \textbf{2024 Revenue (M USD)} & \textbf{Weibo Posts / Members} & \textbf{Reddit Posts / Members} \\
\midrule
\cellcolor{gray}\textbf{\textit{Love and Deepspace}} & \cellcolor{gray}恋与深空 & \cellcolor{gray}722.71 & \cellcolor{gray}146.1M / 4.3M & \cellcolor{gray}98K / 140K \\
\textit{Beyond the World} & 世界之外 & 178.20 & 48.0M / 1.4M & - / - \\
\textit{Light and Night} & 光与夜之恋 & 153.16 & 250.6M / 5.1M & 79 / 443 \\
\textit{Ashes of the Kingdom} & 如鸢（代号鸢） & 77.50 & 184.2M / 1.5M & 26 / 7 \\
\cellcolor{gray}\textbf{\textit{Mr.\ Love: Queen's Choice}} & \cellcolor{gray}恋与制作人 & \cellcolor{gray}40.28 & \cellcolor{gray}176.7M / 2.9M & \cellcolor{gray}11K / 10K \\
\cellcolor{gray}\textbf{\textit{Tears of Themis}} & \cellcolor{gray}未定事件簿 & \cellcolor{gray}39.32 & \cellcolor{gray}118.4M / 2.0M & \cellcolor{gray}13K / 28K \\
\textit{Lovebrush Chronicles} & 时空中的绘旅人 & 29.89 & 131.3M / 4.3M & 1.08K / 3.1K \\
\bottomrule
\end{tabular}
}
\end{table*}
\end{CJK}

\mypara{Contributions} 
Our work makes three main contributions:
First, we present the first large-scale empirical study of toxicity in otome game communities.
By analyzing 620,045 posts collected from Weibo and Reddit, we uncover the prevalence, target groups, temporal dynamics, and linguistic characteristics of toxicity in otome game communities.
These findings provide valuable insights for game developers and platform moderators to better understand and manage the online environments of otome game communities. 
Second, we propose an LLM-driven classifier tailored for detecting toxicity in otome game communities, achieving F1-scores of 0.82 on Weibo and 0.78 on Reddit.
This classifier offers a strong foundation for future mitigation efforts.
We publicly release our dataset of 4,308 manually annotated posts on Hugging Face to support future research on training and evaluating toxicity detectors (Appendix~\ref{appendix:open_science}).
Third, we characterize potential-coordination clusters in otome game communities, highlighting observable patterns that warrant further investigation by moderators.
Despite the substantial user base and market potential of otome games, our study reveals that the efforts to govern toxicity in their communities remain minimal, echoing the longstanding neglect that otome games have faced in the gaming industry.
We call for greater attention to the unique challenges faced by otome game communities.

\section{Preliminaries and Related Work}
\label{section:background}

\begin{figure}[t!]
    \begin{minipage}{0.32\linewidth}
        \centering
        \includegraphics[width=\linewidth]{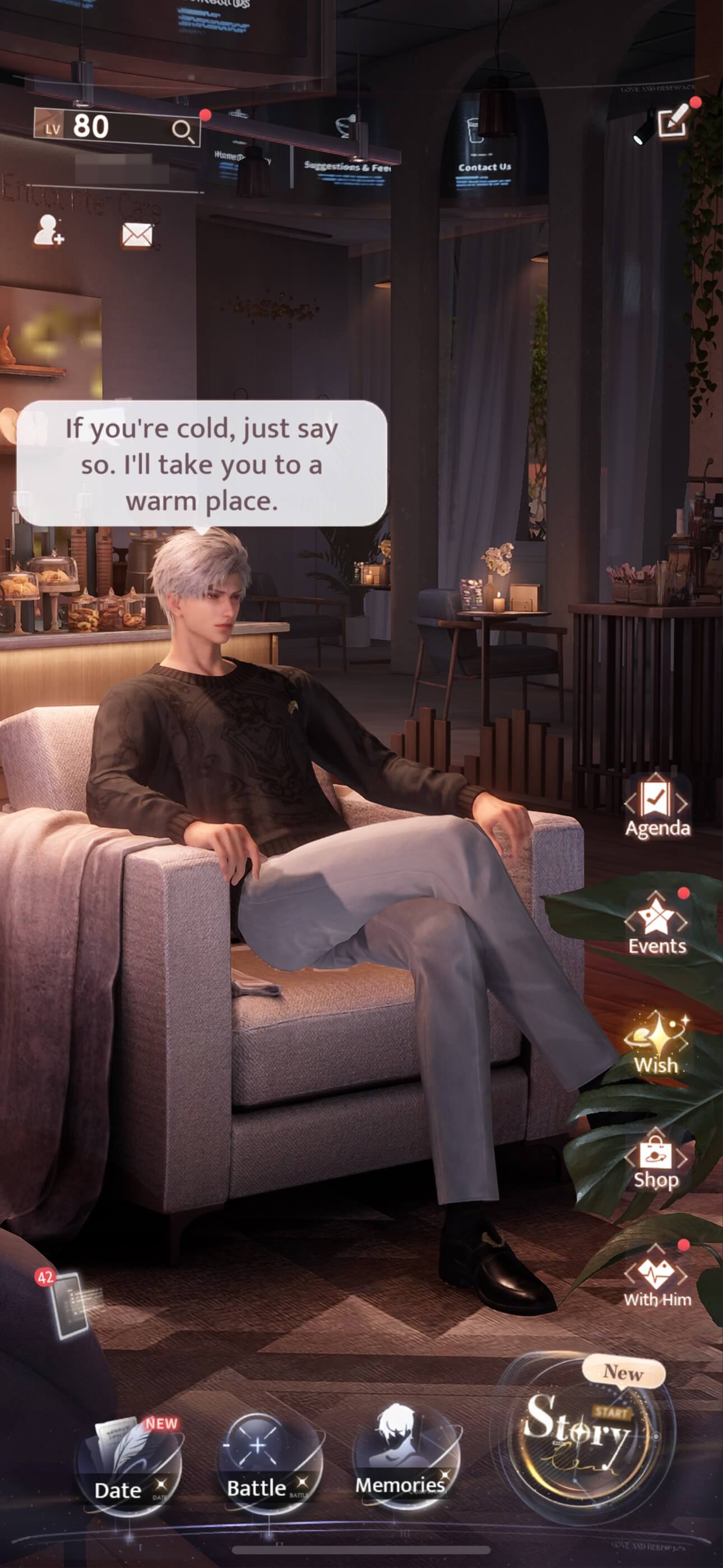}
        \subcaption{Home Screen}
    \end{minipage}
    \hfill
    \begin{minipage}{0.32\linewidth}
        \centering
        \includegraphics[width=\linewidth]{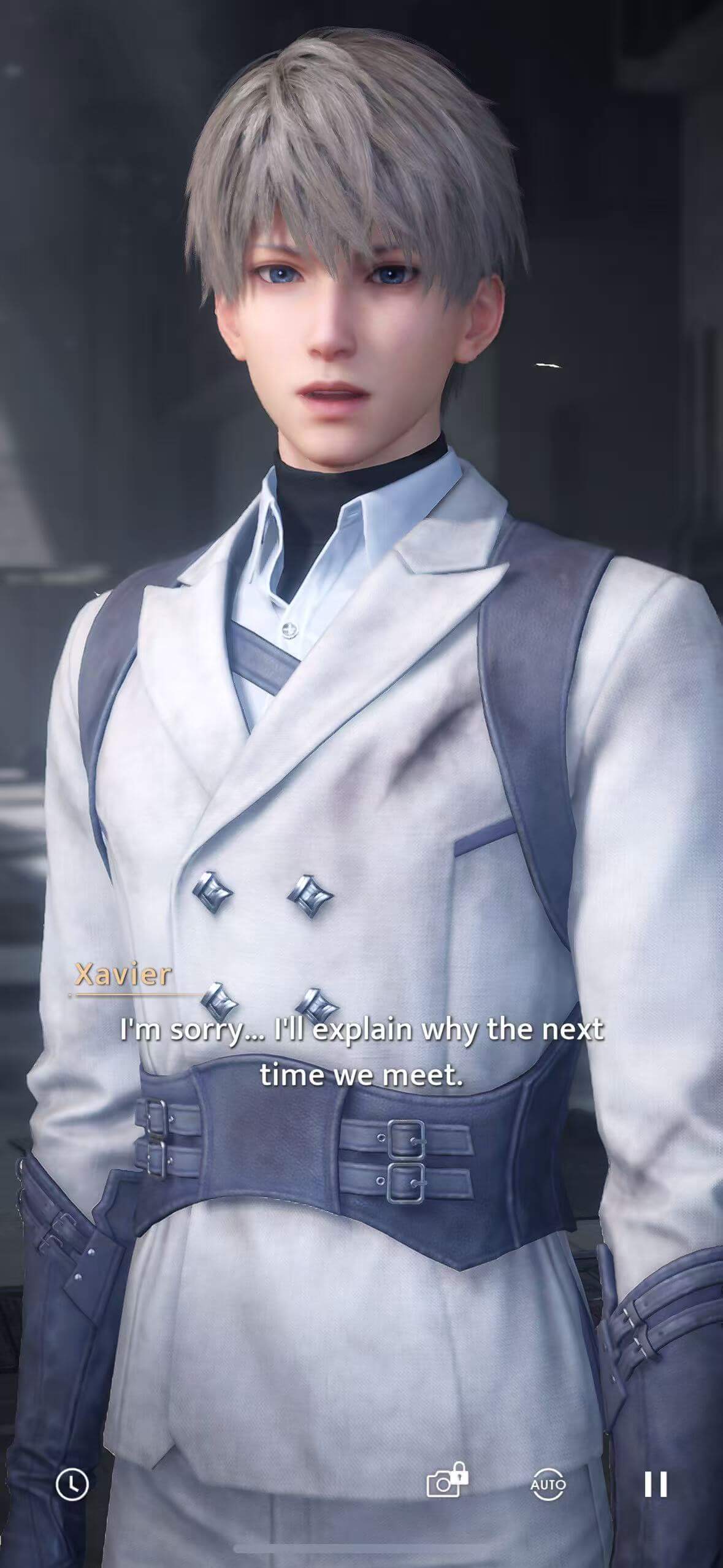}
        \subcaption{Story Mode}
    \end{minipage}
    \begin{minipage}{0.32\linewidth}
    \centering
    \includegraphics[width=\linewidth]{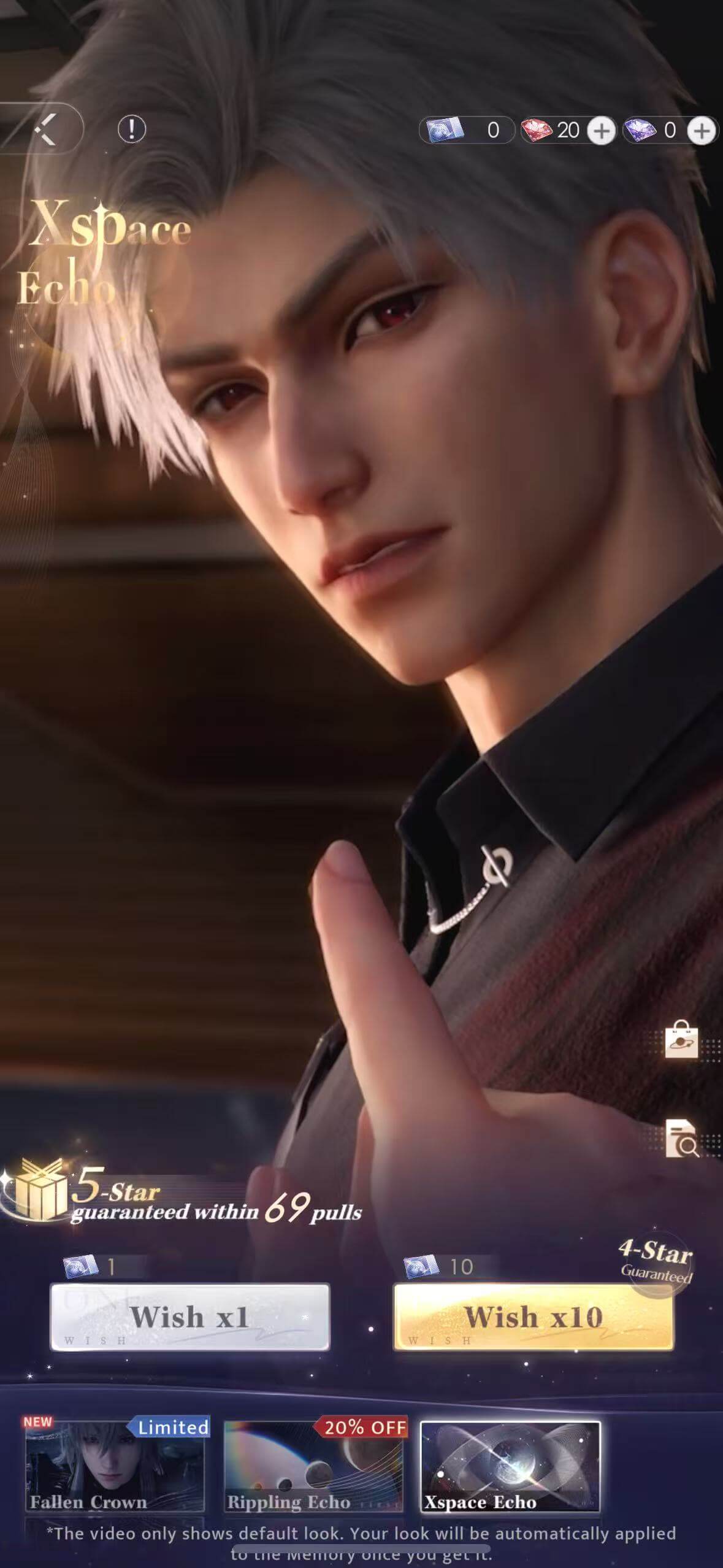}
    \subcaption{Gacha}
    \end{minipage}
    \begin{minipage}{1\linewidth}
    \centering
    \includegraphics[width=\linewidth]{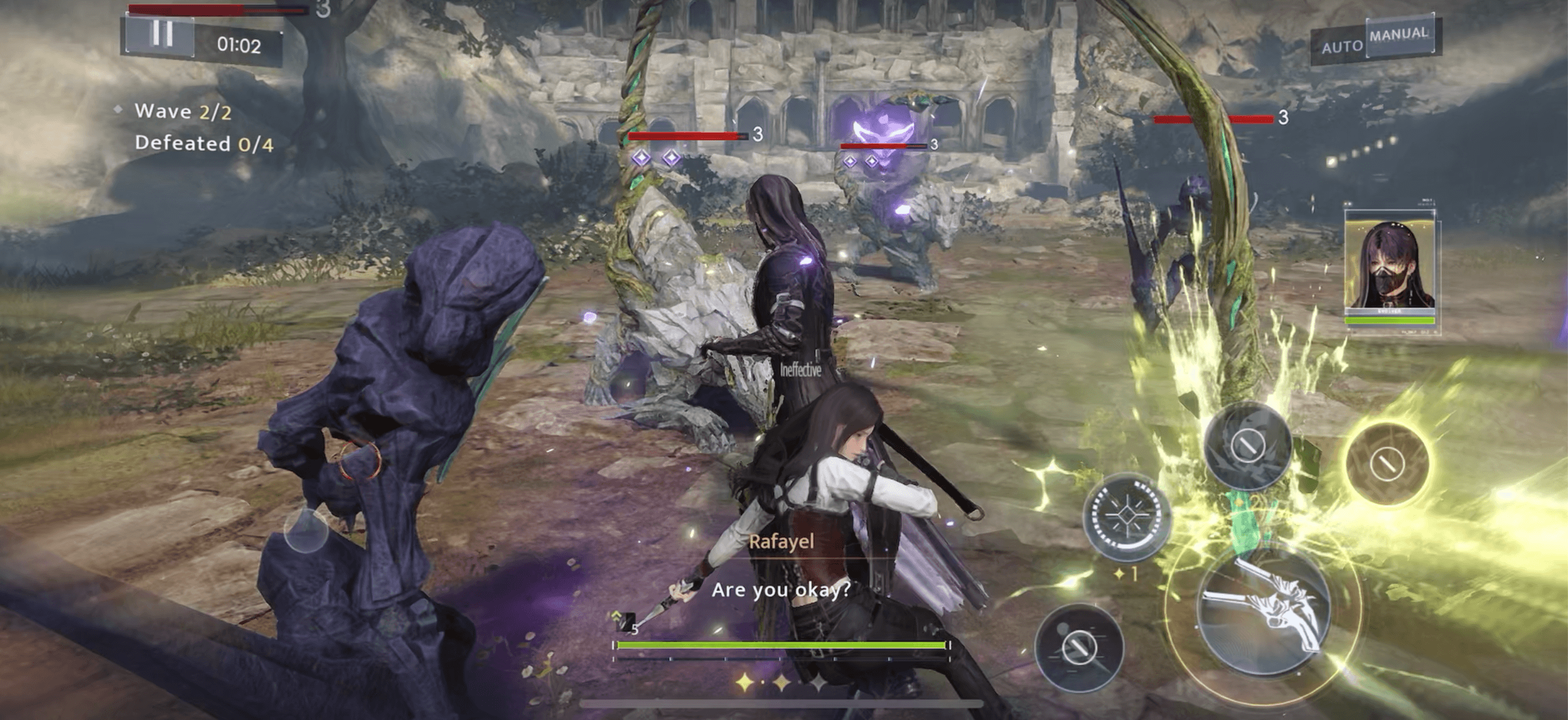}
    \subcaption{Combat Mechanism}
    \end{minipage}
    \caption{Otome game screenshots from the \textit{Love and Deepspace}.
    The game features emotional companionship and fighting alongside male non-player characters.
    }
    \label{figure:LD}
\end{figure}

\mypara{Otome Games}
\emph{Otome game} is a narrative-driven romance simulation genre where the player typically takes on the role of a female protagonist to develop relationships with male non-player characters, through dialogue, choice-based interactions, and combat~\cite{PCFLX25}. 
A representative example is \textit{Love and Deepspace}, as shown in \autoref{figure:LD}, which features a main RPG storyline, card-collecting systems, clue-gathering side episodes, and Live2D cut-ins that deepen immersion.
As illustrated in~\autoref{table:games_overview}, top otome games typically generate hundreds of millions of dollars in annual revenue and attract
millions of followers on social media.
Players congregate in platform-specific communities, such as Weibo's \emph{Super Topics}~\cite{Q23, MGCC21} and subreddits like r/\textit{otomegames}~\cite{RedditOtomeGames}, for discussion and collective action~\cite{DS16}.

Compared to general, mixed-gender gaming communities such as those around \textit{League of Legends} or \textit{Genshin Impact}, otome communities have several structural features that are likely to shape distinct toxicity patterns.
First, the predominantly single-gender player base~\cite{LL24} creates intra-community identity conflicts different from the external gender-based hostility documented in mixed-gender games~\cite{TABBBCDDKKMMRS21}.
Second, the romance-driven design fosters parasocial bonds with characters~\cite{HW56}, turning game updates into emotionally charged community disputes.
Third, fan-circle culture~\cite{MWC23} introduces organized collective behaviors, such as coordinated comment control, voting campaigns, and targeted harassment, that more closely resemble coordinated influence campaigns than general gaming toxicity.
These genre-specific factors motivate a dedicated study rather than direct extrapolation from existing work on general game communities.

Prior work on otome games mainly examines emotional attachment, social support, intimacy, gender, and fan labor.
For example, Lei et al.~\cite{LTHZGT24} explore how players of \textit{Mr.\ Love} seek and provide social support within otome communities.
Other work studies parasocial romantic relationships between female players and male non-player characters~\cite{GH23, GGY25}, the negotiation of female gaze and erotic material under regulatory constraints~\cite{LL24}, fan labor in online otome communities~\cite{G19}, and cosplay commission as a form of commodified or co-created intimacy in the otome community~\cite{ZXZZ24}.
These studies establish otome games as socially and emotionally consequential spaces.
However, the toxicity in otome game communities remains underexplored, such as its prevalence, target groups, and interaction patterns. 
Our work aims to fill this gap.

\mypara{Definition of Toxicity and Toxicity Detection}
In this study, we follow prior work~\cite{JSWSCLBZ24, SSCJJ20} to adopt the Perspective API's definition of toxicity: ``\emph{a rude, disrespectful, or unreasonable comment that is likely to make you leave a discussion}''~\cite{LTTSGMV22}.
For boundary cases involving product or narrative criticism, profanity or strong dissatisfaction alone is not sufficient for a toxicity label; we therefore require that the expression contain direct insults, slurs, threats, or harassment toward a person, group, or character (see~\refappendix{appendix:toxicity_boundary}).

Prior research on gaming communities has extensively documented gender-based hostility, finding that women and LGBTQ+ players face elevated levels of harassment~\cite{TABBBCDDKKMMRS21}, and that underperforming male players can become more hostile toward female players~\cite{KK15}. 
While studies of otome games have explored fan labor and parasocial relationships~\cite{G19, G223, GGY25}, a critical gap remains in quantitatively understanding the toxicity in otome game communities.
This gap further poses a significant challenge to the development of toxicity detection in otome game communities.
While research on large-scale corpora~\cite{WTD17, DWMW17, FDCLBSVSK18} has led to widely-used tools like Perspective API~\cite{LTTSGMV22} and OpenAI Moderation~\cite{OpenAIModerationTooling}, they can be evaded through adversarial attacks like coded speech and homophone substitutions~\cite{HKZP17}. 
The unique, slang-filled discourse of otome communities presents a specific challenge that existing models are ill-equipped to handle, a gap this study aims to address.

\section{\OurMethod}

\begin{figure}[t!]
    \centering
    \includegraphics[width=\linewidth]{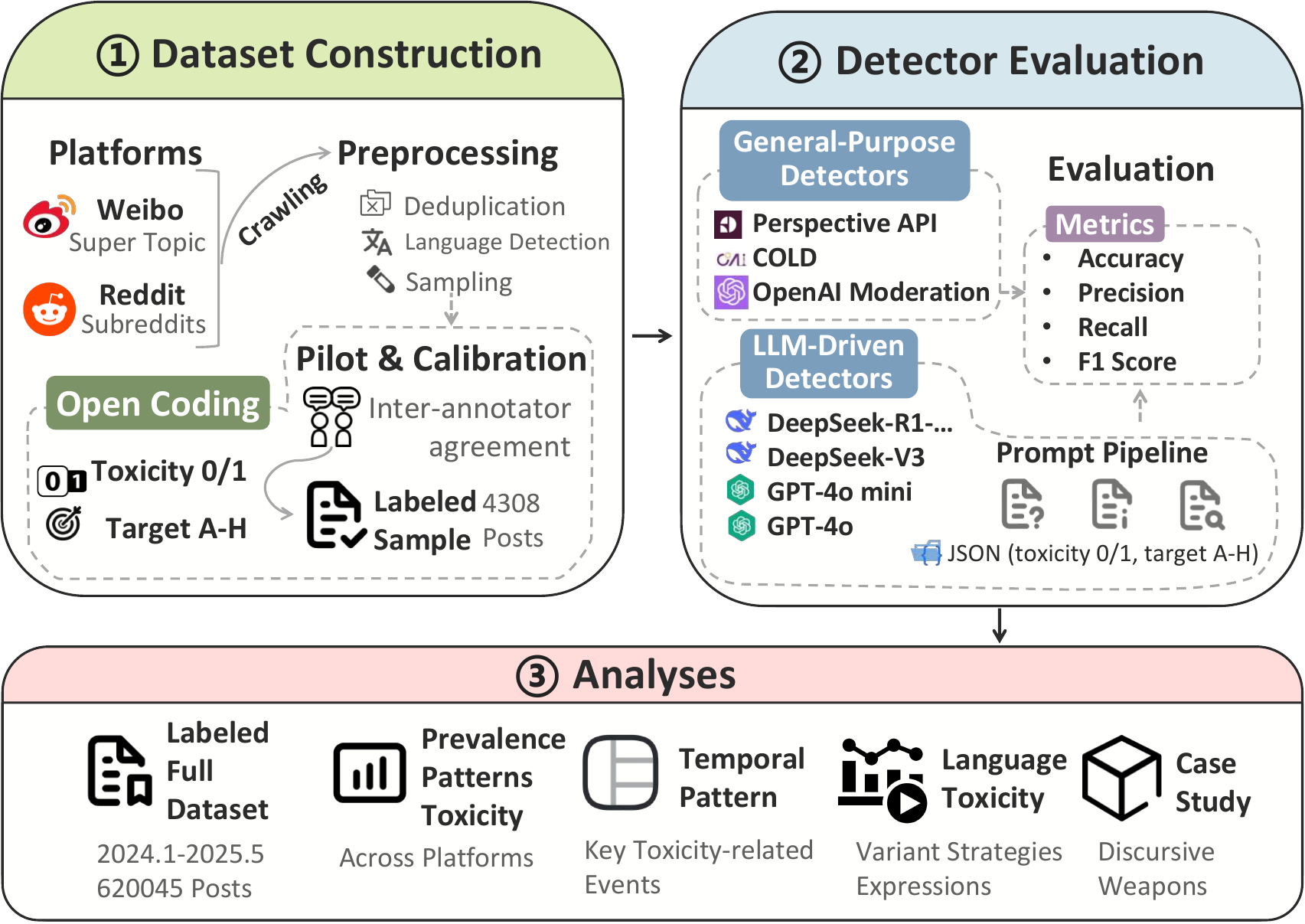}
        \caption{Overview of \OurMethod framework.}
    \label{figure:overview} 
\end{figure}

In this section, we introduce \OurMethod, a framework designed for collecting, evaluating, and analyzing the toxicity in otome game communities.
The overview of \OurMethod is shown in~\autoref{figure:overview}.

\subsection{Dataset Construction}
\label{subsection:dataset_construction}

\subsubsection{Investigated Communities}
\begin{table*}[t!]
\centering
\caption{
The overview of the collected dataset.
}
\label{table:weibo_reddit_combined}
\scalebox{0.80}{
\begin{tabular}{@{}l l| l r r r | l r r r@{}}
\toprule
\multirow{2}{*}{\textbf{Game Type}} & \multirow{2}{*}{\textbf{Community}} & \multicolumn{4}{c|}{\textbf{Weibo}} & \multicolumn{4}{c}{\textbf{Reddit}} \\
\cmidrule(lr){3-10} 
 &  & \textbf{Source} & \textbf{\# Posts} & \textbf{\% Chinese} & \textbf{\# Sample} 
    & \textbf{Source} & \textbf{\# Posts} & \textbf{\% English} & \textbf{\# Sample} \\
\midrule
\multirow{2}{*}{General Games}
  & General Game Community             & Keyword     & 63,295  & 98.09\% & 383 
                        & Subreddit & 78,721  & 94.12\% & 383 \\
  & \textit{Genshin Impact}  & Super Topic & 6,041   & 98.39\% & 363 
                        & Subreddit & 196,943 & 91.12\% & 384 \\
\midrule

\multirow{4}{*}{Otome Games}
  & General Otome Community         & Keyword     & 122,622 & 99.03\% & 384 
                            & Subreddit & 11,980  & 96.72\% & 373 \\
  & \textit{Love And Deepspace}  & Super Topic & 20,328  & 99.16\% & 377 
                            & Subreddit & 89,833  & 96.46\% & 383 \\
  & \textit{Mr. Love}             & Super Topic & 13,563  & 99.85\% & 374 
                            & Subreddit & 419     & 97.14\% & 201 \\
  & \textit{Tears Of Themis}     & Super Topic & 14,031  & 99.74\% & 374 
                            & Subreddit & 2,269   & 93.61\% & 329 \\
\midrule
\multicolumn{2}{l|}{\textbf{Total}} &  & \textbf{239,880} & \textbf{99.04\%} & \textbf{2,255} &  & \textbf{380,165} & \textbf{94.86\%} & \textbf{2,053} \\
\bottomrule
\end{tabular}
}
\end{table*}

We analyze toxicity in four otome game communities, referred to as \textit{study groups}, and compare them with two general game communities, serving as \textit{control groups}.

\mypara{Study Groups}
Our study groups focus on the top three otome games most frequently discussed in Chinese- and English-speaking communities, and general otome community, as shown in~\autoref{table:games_overview}.
\begin{itemize}[leftmargin=*,label=\textbullet]  
    \item \textbf{\textit{Love and Deepspace}} (2024) introduces full-3D models and action gameplay, achieving massive commercial success by attracting over 50 million players in its first year~\cite{WikiLoveAndDeepspace, LoveAndDeepspaceWomenGamers}.
    \item \textbf{\textit{Mr. Love: Queen's Choice}} (2017) is an early adopter of interactive phone-call events, garnering over 10 million downloads in China during its first year~\cite{WikiMrLove}.
    \item \textbf{\textit{Tears of Themis}} (2020) blends courtroom investigation with romance, accumulating over 20 million global installs~\cite{TearsOfThemisGooglePlay}.
    \item \textbf{General Otome Community:} Sourced from \textit{r/otomegames} subreddit and Weibo's ``otome games'' tag, this group reflects a broad and diverse player base of otome games.
\end{itemize}

\mypara{Control Groups}
To distinguish the toxic patterns unique to otome game communities, we include two control groups from broader game communities.
\begin{itemize}[leftmargin=*,label=\textbullet]  
    \item \textbf{\textit{Genshin Impact}} (2020) is an open-world RPG game that attracts a mixed-gender player base~\cite{GenshinRevenueReport}. 
    Although not an otome game, its ongoing story updates and monetization model resemble those of otome games, making it a reasonable game-level comparison.
    \item \textbf{General Game Community:} Aggregated from \textit{r/gaming} on Reddit and the ``Games'' tag on Weibo, this group is used to compare with the general otome community. 
    We regard it as a genre-level comparison.
\end{itemize}

\subsubsection{Data Collection}

We select Reddit and Weibo as the primary studied social platforms for two main reasons.
First, both platforms provide dedicated communities for fan gatherings (e.g., Reddit's subreddits and Weibo's Super Topics), which we denoted as \textit{communities} in this study.
These communities naturally segment fan groups and thus enable us to directly analyze the behaviors of different game fandoms.
Second, these communities are fan-governed, with volunteer moderators on Reddit~\cite{M19} and community hosts on Weibo's Super Topics~\cite{Q23, MGCC21} managing the daily activity. 
They are also closely monitored by game companies, who track them for player feedback and brand management~\cite{G19}.
This dual nature encourages player participation and makes them reliable sources for observing how players negotiate conflict and express grievances.
Specifically, our data collection process is suited to each platform's structure.

\begin{itemize}[leftmargin=*,label=\textbullet]  
    \item \textbf{Reddit:} 
    Reddit is a platform that provides autonomous, volunteer-moderated communities, namely \textit{subreddits}.
    Following prior studies~\cite{CPSGEG17, K20}, we use Arctic Shift\footnote{\url{https://github.com/ArthurHeitmann/arctic_shift}.} an open-source crawler to collect all posts from the subreddits of the study and control groups, that is \textit{r/LoveAndDeepspace}~\cite{RedditLoveAndDeepspace}, \textit{r/MrLove}~\cite{RedditMrLove}, \textit{r/TearsOfThemis}~\cite{RedditTearsOfThemis}, \textit{r/Genshin\_Impact}~\cite{RedditGenshinImpact}, \textit{r/otomegames}~\cite{RedditOtomeGames} and \textit{r/gaming}~\cite{RedditGaming}.

    \item \textbf{Weibo:}
    Similar to Reddit's subreddits, Weibo, one of the largest Chinese-speaking social media platforms, also holds fandom-based communities called \textit{Super Topics}.
    Our retrieval strategy is twofold. 
    For specific games, we collect all posts from their dedicated Super Topics, that is \textit{Love and Deepspace}~\cite{WeiboLoveAndDeepspace}, \textit{Mr. Love: Queen's Choice}~\cite{WeiboMrLove}, \textit{Tears of Themis}~\cite{WeiboTearsOfThemis}, and \textit{Genshin Impact}~\cite{WeiboGenshinImpact}.
    For broader groups such as the general otome community and the general game community, which lack official Super Topics, we instead rely on keyword searches to ensure comprehensive coverage.
    All Weibo data is collect using the \texttt{weibo-search} tool,\footnote{\url{https://github.com/dataabc/weibo-search}.} with the detailed keyword search strategy described in~\refappendix{appendix:keyword_filter}.
\end{itemize}

\subsubsection{Data Pre-Processing}

\mypara{Deduplication}
To ensure that each sample in the dataset represents a unique post, we perform a deduplication process.
We identify and remove any duplicate posts based on their unique post identifiers (\texttt{ID}) assigned by the platform.
We also exclude posts marked as deleted or removed on either platform during this stage (see~\refappendix{appendix:ethics}).

\mypara{Language Verification}
Although Reddit's subreddits and Weibo's Super Topics analyzed in this study are typically considered English- and Chinese-speaking communities, respectively, we verified their language distributions through the \texttt{lingua} toolkit.\footnote{\url{https://github.com/pemistahl/lingua}.}
Our results show that 94.86\% of Reddit posts are in English, while 99.04\% of Weibo posts are in Chinese, as illustrated in~\autoref{table:weibo_reddit_combined}. 
Given this high degree of monolingualism, we henceforth treat them as English-speaking and Chinese-speaking communities in our analysis.
We acknowledge this choice may exclude content in other languages, and we discuss this further in~\autoref{section:discussion}.

\mypara{Data Statistics}
In total, we collect 239,880 posts from Weibo and 380,165 posts from Reddit, which, to the best of our knowledge, corresponds to the largest dataset to date on community discussions of otome games. 
For each post, we collect its ID, content, creation time, number of comments, tags, and user IDs.  
Our data collection spans from January 1, 2024, to May 31, 2025. 
This time frame allows us to capture the initial growth phase of a newly launched game (\textit{Love and Deepspace}, released on January 18, 2024) and multiple content update cycles of other established games.
\autoref{table:weibo_reddit_combined} summarizes the statistics of posts across platforms and communities, including sample sizes and language distributions.
Note that only 419 posts are collected from the \textit{Mr. Love} subreddit.
This is because, by 2024, the game was already entered its eighth year: while its Chinese-speaking community remains active, the English-speaking market begins to contract, resulting in a decline in user postings.
Nevertheless, we include this community in our dataset because it provides valuable and irreplaceable perspectives on community dynamics in the late stages of an otome game's lifecycle, as later discussed in~\autoref{section:temporal}.

\begin{table*}[t!]
\centering
\caption{Codebook of toxicity labels and target groups. We paraphrased the examples to prevent verbatim searches from identifying the users while preserving the original meaning.}
\label{table:toxicity_target_codebook}
\tabcolsep 2pt
\scalebox{0.8}{
\begin{tabular}{@{}ccl>{\arraybackslash}p{5.5cm}>{\arraybackslash}p{6.5cm}rr@{}}
\toprule
\textbf{Task} & \textbf{NO.} & \textbf{Code} & \textbf{Description} & \textbf{Example} & \textbf{\% Weibo}& \textbf{\% Reddit} \\
\midrule
\multirow{2}{*}{\textbf{Toxicity}}
 & 0 & non-toxic & Language not likely to drive others away & \texttt{This update offers too little content for its price.}  & 81.48\% & 95.50\%\\
 & 1 & toxic & Language likely to drive others away & \texttt{Everyone in this fandom is a pathetic loser, just shut up already.}  & 18.52\% & 4.50\%\\
\midrule
\multirow{8}{*}{\textbf{\shortstack{Target\\Groups}}}
 & A & players & Game players and communities & \texttt{Otome fans are clueless idiots who turn every thread into a fight.} & 43.33\% & 7.71\%\\
 & B & NPCs & In-game characters like non-player characters (NPCs) and main character (MC) & \texttt{That love interest is a disgusting loser, and I'm sick of seeing him.} & 5.96\% & 33.89\%\\
 & C & game developers & Game content, developers, designers, and marketing & \texttt{The devs are greedy morons who treat players like wallets.} & 34.19\% & 22.10\%\\
 & D & platform moderators & Moderators or event organizers of the social platform & \texttt{The mods here are useless bullies who just abuse their authority.} & 1.88\% & 2.34\%\\
 & E & policymakers & Censorship bodies or government regulators & \texttt{The regulators who wrote these rules are brainless fools.} & 0.22\% & 0.64\%\\
 & F & identity groups & men, women, LGBTQ+, social classes & \texttt{Women are too stupid to understand game design and should stay quiet.} & 7.76\% & 20.00\%\\
 & G & other game-related entities & Other game-related entities like voice actors, cosplayers, co-branding brands & \texttt{That voice actor is talentless trash and should just quit.} & 3.89\% & 0.64\%\\
 & H & unknown & Unclear, mixed, or sarcastic target & \texttt{What an insufferable clown. Absolutely useless.} & 2.76\% & 12.69\%\\
\bottomrule
\end{tabular}
}
\end{table*}

\subsubsection{Data Sampling and Human Annotation}
\label{section:annotation_process}

Given that toxicity in game communities remains largely unquantified, we begin our analysis by sampling data from the collected dataset.
Notably, the sampled dataset includes both otome and general game communities, as we aim to provide a systematic comparison among them. 
We then manually annotate two types of labels: toxicity and target groups, to gain deeper insights into the nature of toxicity within these communities.

\mypara{Sampling}
Following previous studies~\cite{C77, L19}, we calculate the minimum required sample size for each community to balance manageability with statistical representativeness.
Specifically, we first apply the standard formula for estimating a population proportion~\cite{estimate_population}, then adjust the result using the finite population correction (FPC)~\cite{estimate_population, L19} to account for the specific size of each community (details in~\refappendix{appendix:sampling_stats}).
The final sample sizes are summarized in~\autoref{table:weibo_reddit_combined}, including 2,255 Weibo posts and 2,053 Reddit posts.

We then perform human annotation to establish ground-truth labels for subsequent evaluation and analysis.
The annotation process is designed to produce two levels of labels:
\textit{1) Toxicity:} each sample is labeled as either toxic or non-toxic, based on the definition of toxicity outlined in~\autoref{section:background}. 
These labels serve as the ground truth for evaluating toxicity detectors in~\autoref{section:detector_evaluation};
\textit{2) Target groups:} for samples identified as toxic, annotators employ open coding to identify the specific groups targeted.
This enables a more fine-grained investigation of the affected groups in otome game communities.
To avoid conflating disagreement with toxicity, we apply explicit boundary rules during annotation.
Representative boundary cases are provided in~\refappendix{appendix:toxicity_boundary}.

To ensure both rigor and domain relevance, we structure our annotation process in two phases: a pilot study to develop and calibrate the annotation schema, followed by full-scale annotation of the sampled set.
The annotation is led by two expert annotators with over six years of experience as otome game players.

\noindent\textit{\underline{\textit{Pilot Study.}}}
We first sample 578 posts from the sample set to conduct a pilot study.
In this phase, two annotators independently labeled each post for toxicity and performed open coding to identify the target groups referenced in the toxic posts.
The Cohen's Kappa score of the toxicity label is 0.87.
The annotators then work together to develop a codebook of the target groups.
In the end, they identify eight target groups and re-code the pilot data to ensure consistency.
Throughout the pilot phase, inter-annotator agreement for the target group label steadily improves, with Cohen's Kappa increasing from an initial 0.40 to 0.84, indicating a substantial improvement in annotation reliability.

\noindent\textit{\underline{\textit{Full Annotation.}}}
With the finalized codebook, the two annotators independently label the remaining sampled posts and resolve disagreements through discussion.
No new target groups are identified in this phase.
The annotation achieves a Cohen's Kappa of 0.84 for the binary \texttt{Toxicity} label and 0.82 for the eight-category \texttt{Target Group} label, indicating a high level of consistency between annotators.
Note, we observe fewer than five posts that target multiple groups within a single sample during annotation.
In such cases, the annotators label the group subjected to the most severe abuse.
The codebook is available in~\autoref{table:toxicity_target_codebook}, with additional examples provided in~\autoref{table:toxicity_examples} in the Appendix.

\begin{table*}[t!]
\centering
\caption{Binary toxicity detection performance.}
\label{table:toxicity_binary}
\tabcolsep 3pt
\scalebox{0.77}{
\renewcommand{\arraystretch}{1.15}
\begin{tabular}{@{}l|ccccccc|ccccccc@{}}
\toprule
\multirow{2}{*}{\textbf{Model}} &
\multicolumn{7}{c|}{\textbf{Weibo}} &
\multicolumn{7}{c}{\textbf{Reddit}} \\
\cmidrule(lr){2-8}\cmidrule(lr){9-15}
& ACC & Prec. & Recall & F1 & micro-F1 & macro-F1 & w-F1
& ACC & Prec. & Recall & F1 & micro-F1 & macro-F1 & w-F1 \\
\midrule
Perspective API
& 0.85 & 0.37 & 0.58 & 0.45 & 0.85 & 0.68 & 0.86
& 0.93 & 0.25 & 0.39 & 0.30 & 0.93 & 0.63 & 0.94 \\
COLD
& 0.91 & 0.81 & 0.20 & 0.33 & 0.91 & 0.64 & 0.88
& 0.95 & 0.00 & 0.00 & 0.00 & 0.95 & 0.49 & 0.94 \\
OpenAI Moderation API
& 0.89 & 0.41 & 0.12 & 0.18 & 0.89 & 0.56 & 0.86
& 0.95 & 0.17 & 0.06 & 0.09 & 0.95 & 0.53 & 0.94 \\
\midrule
LLM-Driven (DeepSeek-R1-Distill-Qwen-14B)
& 0.51 & 0.09 & 0.40 & 0.15 & 0.51 & 0.40 & 0.60
& 0.63 & 0.04 & 0.37 & 0.08 & 0.63 & 0.42 & 0.74 \\
LLM-Driven (DeepSeek-V3)
& \textbf{0.96} & 0.84 & \textbf{0.80} & \textbf{0.82} & \textbf{0.96} & \textbf{0.90} & \textbf{0.96}
& 0.98 & 0.74 & 0.57 & 0.64 & 0.98 & 0.82 & 0.98 \\
LLM-Driven (GPT-4o mini)
& 0.95 & 0.79 & 0.72 & 0.75 & 0.95 & 0.86 & 0.95
& \textbf{0.99} & 0.78 & \textbf{0.78} & \textbf{0.78} & \textbf{0.99} & \textbf{0.89} & \textbf{0.99} \\
LLM-Driven (GPT-4o)
& 0.93 & \textbf{0.88} & 0.40 & 0.55 & 0.93 & 0.76 & 0.92
& 0.97 & \textbf{0.90} & 0.10 & 0.18 & 0.97 & 0.58 & 0.96 \\
\bottomrule
\end{tabular}
}
\end{table*}

\begin{table*}[t!]
\centering
\caption{Target group identification performance.}
\label{table:target_group}
\scalebox{0.77}{
\renewcommand{\arraystretch}{1.15}
\begin{tabular}{@{}l|cccc|cccc@{}}
\toprule
\multirow{2}{*}{\textbf{Model}} &
\multicolumn{4}{c|}{\textbf{Weibo}} &
\multicolumn{4}{c}{\textbf{Reddit}} \\
\cmidrule(lr){2-5}\cmidrule(lr){6-9}
& ACC & micro-F1 & macro-F1 & w-F1
& ACC & micro-F1 & macro-F1 & w-F1 \\
\midrule
DeepSeek-R1-Distill-Qwen-14B
& 0.15 & 0.15 & 0.07 & 0.14
& 0.22 & 0.22 & 0.16 & 0.21 \\
DeepSeek-V3
& \textbf{0.89} & \textbf{0.89} & \textbf{0.80} & \textbf{0.88}
& 0.74 & 0.74 & 0.66 & 0.73 \\
GPT-4o mini
& 0.82 & 0.82 & 0.72 & 0.81
& \textbf{0.85} & \textbf{0.85} & \textbf{0.79} & \textbf{0.84} \\
GPT-4o
& 0.60 & 0.60 & 0.46 & 0.58
& 0.82 & 0.82 & 0.76 & 0.81 \\
\bottomrule
\end{tabular}
}
\end{table*}

\subsection{Detector Evaluation}
\label{section:detector_evaluation}

We evaluate existing general-purpose detectors and our proposed LLM-driven detectors on the ground truth dataset.
We begin by introducing the two model families under evaluation, followed by a description of the experimental settings and results.
Finally, we identify the most effective models for subsequent analysis.

\subsubsection{Detection Models}
\label{section:models}
We evaluate two distinct families of models: general-purpose and LLM-Driven detectors.

\mypara{General-Purpose Detectors}
We benchmark three widely used detectors to establish a baseline. 
These models represent well-established tools designed to detect general-purpose toxicity (rather than otome-specific toxicity), which are the Perspective API~\cite{LTTSGMV22}, the OpenAI Moderation API~\cite{OpenAIModerationTooling, O24}, and the Chinese-specific COLD classifier~\cite{DZSZMMH22}. 

\mypara{LLM-Driven Detectors}
Recognizing the unique context of otome game communities, we also design and evaluate several LLM-driven detectors. 
To determine the optimal configuration, we performed ablation studies on prompt design: comparing Chain-of-Thought (CoT), Definition-only, and Reasoning prompts and the number of examples in context ($N$, from 0 to 15), as detailed in~\refappendix{appendix:prompt_optimization}. 
Our findings consistently show that a 5-shot Reasoning prompt achieves the best balance of performance and cost, yielding the highest F1-scores on both Weibo (0.82) and Reddit (0.78). 
Consequently, we adopt this configuration for all subsequent experiments.
We use in-context prompting rather than supervised fine-tuning because our annotated set is primarily intended for validation, and fine-tuning on a small domain-specific sample risks overfitting to the sampled communities and events.
Since the 5-shot Reasoning prompt already substantially outperforms general-purpose detectors, we adopt it as a practical and reproducible detector for this first measurement study, while leaving supervised fine-tuning to future work.
For the backend, we test four representative LLMs: DeepSeek-V3~\cite{deepseekv3}, DeepSeek-R1-Distill-Qwen-14B~\cite{DeepSeekR1DistillQwen14B}, GPT-4o~\cite{GPT4o}, and GPT-4o mini~\cite{GPT-4o-mini}. 
Further details of these models and prompt design are available in~\refappendix{appendix:models_details} and~\refappendix{appendix:prompt}.

\subsubsection{Experimental Settings}
\label{section:experiments}

We detail the specific settings for the seven models evaluated in our study.
For the Perspective API, we use the toxicity attribute. 
Since it does not provide official thresholds, we determined the optimal thresholds that maximize the F1-score on the ground truth dataset, setting them to 0.28 for Weibo and 0.20 for Reddit. 
For the OpenAI Moderation API, we adopt the overall flagged label returned by the omni-moderation-latest model, which indicates whether a text violates any of the covered categories (e.g., hate, harassment, self-harm).
For COLD, we use the publicly available pre-trained checkpoints without additional fine-tuning.
For the LLM-driven detectors, we set the temperature to 1.0, which balances consistency with the need for nuanced language understanding, as required for our analysis task~\cite{DeepSeekTemperature}.
All other hyperparameters are left at their default settings. 
The backend endpoint of each detector is DeepSeek-V3 (0324),\footnote{\url{https://platform.deepseek.com/}.} 
DeepSeek-R1-Distill-Qwen-14B,\footnote{\url{https://huggingface.co/deepseek-ai/DeepSeek-R1-Distill-Qwen-14B}.} 
GPT-4o (2024-08-06),\footnote{\url{https://platform.openai.com/docs/models/gpt-4o}.} 
and GPT-4o mini (2024-07-18).\footnote{\url{https://platform.openai.com/docs/models/gpt-4o-mini}.}
Following standard practice~\cite{SSBZ25, JSWSCLBZ24}, we report Accuracy, Precision, Recall, and macro F1-score for the toxicity detection task. 
Given the class imbalance, we focus on Recall and F1 as they provide a fairer assessment of minority-class performance~\cite{SR15}. 
For the eight-class target group identification task, we report Accuracy as the primary metric.

\subsubsection{Experimental Results and Model Selection}
\label{section:evaluation}
The binary toxicity detection performance is reported in~\autoref{table:toxicity_binary}, and the target group identification performance of the LLM-driven detectors is reported in~\autoref{table:target_group}. 

\mypara{Binary Toxicity Detection} 
For toxicity detection, we find that general-purpose detectors (Perspective, COLD, OpenAI Moderation) show poor performance, achieving F1 scores of 0.45, 0.33, and 0.18 on Weibo, respectively. 
These models exhibit extreme precision–recall imbalance or fail when applied outside their specific training language.
In contrast, LLM-driven detectors perform substantially better, particularly when their primary training data aligns with the platform's dominant language.
DeepSeek-V3, for example, whose training data includes a substantial amount of Chinese text, reaches the highest F1-score (0.82) on Chinese Weibo, while GPT-4o-mini leads on English Reddit with 0.78.
These results demonstrate that LLMs, with their ability to capture nuanced and context-dependent expressions, are far better suited for toxicity detection in the otome game context.
In addition, LLM-driven detectors offer better explainability compared to general-purpose detectors. 
By prompting the model to output the toxicity reason, we are able to perform more fine-grained analysis of toxic posts, such as identifying toxic spans, as demonstrated in~\autoref{section:Linguistic-Strategies}.

\mypara{Target Group Identification} 
As shown in~\autoref{table:target_group}, LLM-driven detectors again demonstrate the importance of language alignment for target group identification. 
DeepSeek-V3 achieves the highest accuracy on Weibo (0.89), while GPT-4o-mini performs best on Reddit (0.85).
General-purpose detectors do not support target group classification, so we omit them from this analysis.

\mypara{Model Selection} 
Based on above analysis, we select DeepSeek-V3 for Weibo and GPT-4o-mini for Reddit for our large-scale analysis. 
We acknowledge that none of the models is perfect; therefore, we provide a detailed error analysis below~\refappendix{appendix:error}.

\subsection{Analysis}

Using the best performing LLM-driven detectors identified by the \OurMethod, we label the full dataset of 620,045 posts.
The following sections analyze this comprehensive dataset to investigate the prevalence and patterns of toxicity (\autoref{section:RQ1}), its temporal dynamics (\autoref{section:temporal}), and its linguistic expression (\autoref{section:Linguistic-Strategies}).

\section{Prevalence and Patterns of Toxicity}
\label{section:RQ1}

In this section, we address \textbf{RQ1} by comparing the prevalence, interaction patterns, target groups, and thematic content of toxic posts across platforms.
We first examine toxicity prevalence and interaction patterns, then analyze target groups, and finally compare the themes of toxic disputes.

\begin{figure}[t!]
    \centering
    \includegraphics[width=\linewidth]{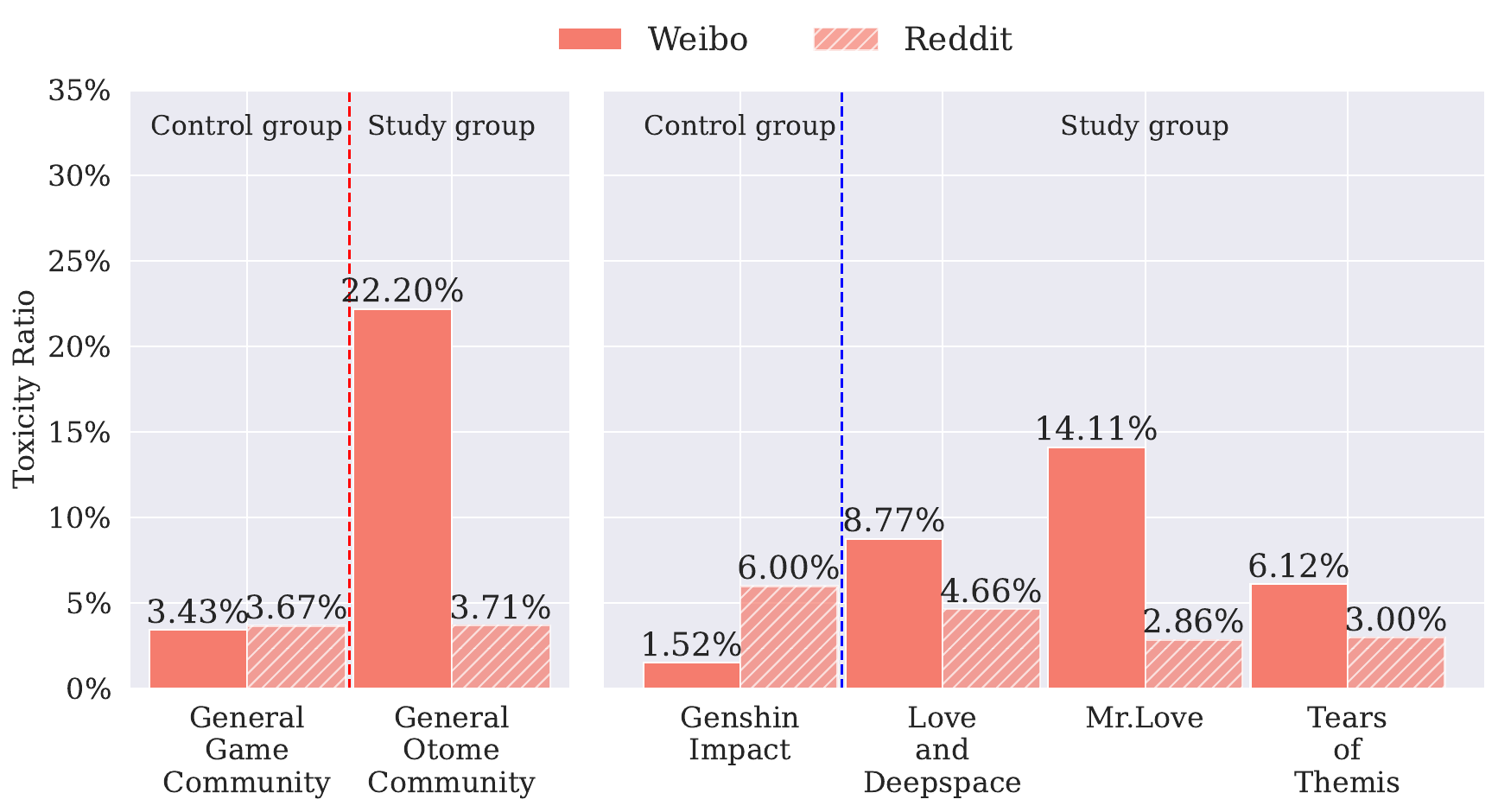}
    \caption{Toxicity ratios across platforms and otome game communities.
    }
    \label{figure:toxicity}
\end{figure}

\mypara{Overall Toxicity Analysis}
As shown in~\autoref{figure:toxicity}, the toxicity patterns between Weibo and Reddit exhibit significant differences.
While the toxicity ratios for the general game community are similar on both platforms (3.43\% on Weibo vs. 3.67\% on Reddit), the introduction of otome game-specific content reveals a dramatic divergence. 
On Weibo, the toxicity ratio in the general otome community surges to 22.20\%, a more than sixfold increase over the general game community, suggesting that otome-related discussions may be particularly prone to toxic expression in Chinese-language platforms.
In contrast, the toxicity ratio in Reddit's otome communities remains relatively low at 3.71\%, closely aligned with the platform's general gaming toxicity levels.
This gap is large and precisely estimated (Weibo 22.20\%, Wilson 95\% CI [21.9, 22.4], vs.\ Reddit 3.71\%, CI [3.4, 4.1], relative risk $=5.99$, Cohen's $h=0.59$), whereas the two control communities show negligible or reversed cross-platform differences (general games $h=-0.01$, \textit{Genshin Impact} $h=-0.25$), suggesting that the gap is specific to otome communities rather than an artifact of cross-platform measurement.
Besides, individual games demonstrate distinct toxicity patterns across platforms and genres. 
On Weibo, the three otome games, \textit{Mr. Love} (14.11\%), \textit{Love and Deepspace} (8.77\%), and \textit{Tears of Themis} (6.12\%), all exhibit higher toxicity ratios than the control game \textit{Genshin Impact} (1.52\%). 
In contrast, on Reddit, the toxicity of the otome games is consistently low, ranging from 2.86\% to 4.66\%, and remains below the toxicity ratio of \textit{Genshin Impact} (6.00\%).
This divergence suggests that on Weibo, toxicity may be shaped more by the specific characteristics of the otome game genre, whereas on Reddit, it seems to be influenced more by the platform-wide cultural norms.

\mypara{Impacts of Toxicity on User Interaction}
User interaction patterns, such as likes and comments, can reveal whether a platform's socio-technical ecosystem algorithmically amplifies or suppresses toxic content~\cite{MCWPZD23}.
Therefore, to understand how different platform environments shape community engagement with toxic posts, we investigate the relationship between user engagement and toxic posts by fitting a logistic regression model for each platform-community pair.
The resulting standardized coefficients and their significance are reported in~\autoref{table:regression_interactions}.
Interestingly, we find that the correlations between a post's toxicity and its user engagement on Weibo and Reddit are opposing.
On Weibo, we observe a consistent negative correlation: toxic posts typically receive fewer likes and comments.
For instance, in the \textit{Love and Deepspace} community, the comment coefficient is -2.01 ($p<.001$). 
Conversely, on Reddit, toxic posts are positively correlated with comment counts, particularly in the \textit{Mr.~Love} subreddit, where we find a strong positive effect ($\beta_2 = +0.784$, $p < .001$).
This suggests that, unlike on Weibo, toxic content on Reddit tends to provoke interaction rather than suppress it.
However, this suppression effect on Weibo is not uniform.
Among the 24{,}311 toxic Weibo posts, those employing linguistic evasion strategies such as homophone substitution and slang ($n$\,=\,16{,}700) receive significantly more comments (mean 26.76 vs.\ 8.37, $p<10^{-40}$) and reposts (mean 24.64 vs.\ 10.98, $p<10^{-39}$) than those without, suggesting that Weibo's suppression primarily operates at the visibility level, while linguistic evasion may circumvent these mechanisms.
These opposing patterns may reflect two distinct socio-technical governance regimes.
The negative correlation on Weibo could suggest a platform-level discouragement of posting toxic content.
This aligns with prior research on Chinese social media, which finds that both state censorship and platform-specific norms act to silence collective expression and discourage public conflict, effectively marginalizing such content~\cite{KPR13, R18}. 
In contrast, the positive correlation on Reddit aligns with the platform's engagement-driven design~\cite{GPW17, MAT13}.
In such settings, interactions, including toxic speech and counter speech, play a critical role in shaping content visibility~\cite{M19}.

\begin{figure}[t!]
    \centering
    \includegraphics[width=\linewidth]{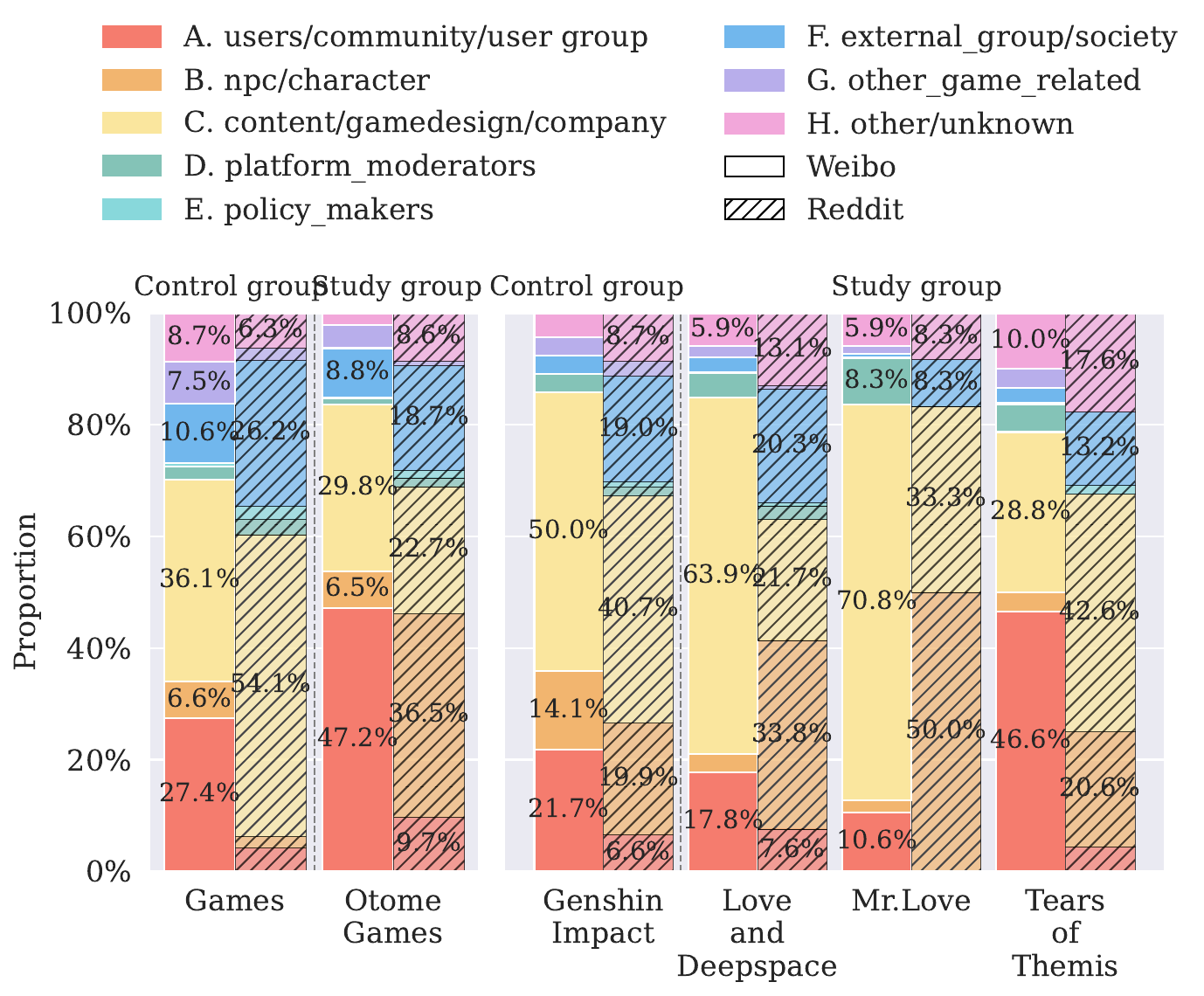}
    \caption{Proportion of target groups for Weibo and Reddit toxic posts.
    }
    \label{figure:target}
\end{figure}

\mypara{Target Group Analysis}
To compare whom toxic posts target across platforms, we analyze the distribution of target groups shown in~\autoref{figure:target}.
We find that the primary target groups of toxic posts differ notably across platforms.
The platform difference in target-group distribution is statistically significant ($\chi^2$ test, $p<10^{-180}$).
On Weibo, players (Target A) and game developers (Target C) are the most frequently targeted groups, accounting for  47.20\% and 29.84\% of toxic posts in the general otome community, respectively.
In contrast, on Reddit, NPCs (Target B) and game developers (Target C) are more commonly targeted, while toxic posts directed at players (Target A) are relatively rare. 
Specifically, in the general otome community, 36.49\% of toxic posts target NPCs and 22.75\% target game developers, compared to just 9.68\% directed at players.
Regarding individual games, the distribution of targeted groups varies significantly. 
While \textit{Tears of Themis} aligns with the toxicity pattern observed in the general otome community, i.e., primarily targeting players (Target A, 46.57\%), \textit{Mr. Love} and \textit{Love and Deepspace} direct 70.81\% and 63.94\% toxic posts on game developers (Target C), respectively.
These proportions even exceed in the \textit{Genshin Impact} control group (50.00\%).
This suggests that the toxicity in Weibo otome communities is highly heterogeneous and game-dependent.
Conversely, the pattern on Reddit is more consistent. 
NPCs (Target B) are consistently the most frequently targeted group, especially in \textit{Mr. Love} (50.00\%) and \textit{Love and Deepspace} (33.79\%). 
This suggests that when players' strong parasocial attachments are met with narrative frustration, the resulting toxic outbursts are directed at the characters themselves~\cite{HW56, G05, W15}.
This NPC-focused pattern is notably absent in the \textit{Genshin Impact} control group, where game developers remain the primary target, suggesting it is genre-specific.
Although fictional characters cannot themselves be harmed, abusive attacks on them may affect real users who form strong parasocial attachments to these characters and rely on otome
communities for social support~\cite{GGY25, LTHZGT24}.
Such attacks may be perceived as hostility toward users' preferences or community identity, provoke interpersonal conflict, and make community discussions less welcoming. 
We therefore interpret NPC-directed toxicity as a potential risk to community interaction.
We conduct several robustness checks to validate these findings, including bootstrap confidence intervals, confusion-invariant analysis, an NPC-vs-developer comparison, and broader-community sanity checks against non-otome fandom and gaming communities.
Full details are provided in~\refappendix{appendix:rq1_validation}.

\mypara{Thematic Analysis of Toxic Posts}
To compare what toxic disputes concern, we manually code samples of 379 Weibo and 356 Reddit toxic posts selected using the same sampling procedure as our main annotation.
Each sample size corresponds to a $\pm$5\% margin of error at 95\% confidence.
Factional attacks or identity-marking language appear in 55.4\% of the coded Weibo posts, compared with 9.8\% on Reddit, whereas 51.0\% of the coded Reddit posts concern parasocial or consumer/developer grievances.
Moreover, 42.4\% of the coded Weibo posts use in-group slang, a pattern consistent with identity signaling but not sufficient to establish users' motivations.
Overall, the coded Weibo toxicity more often centers on factional conflict, whereas the coded Reddit toxicity more often concerns characters, game content, or developers.

\begin{tcolorbox}[colback=gray!25!white, size=title,breakable,boxsep=1mm,colframe=white,before={\vskip1mm}, after={\vskip0mm}]
\textbf{Take-Aways:}
Toxicity in otome game communities differs from that in general game communities in both prevalence and target structure. 
In general game communities, toxicity is directed at game content, design, companies, or operations.
Otome communities, by contrast, demonstrate community- and narrative-specific targets:
The toxicity levels on Weibo are significantly higher, with player being particularly targeted; on Reddit, hostility shifts toward NPCs and fictional characters.
These patterns differ from those in the general gaming communities we sampled, although similar patterns may exist in subcommunities we did not cover.
\end{tcolorbox}

\section{Temporal Dynamics}
\label{section:temporal}

To address \textbf{RQ2}, this section investigates the temporal dynamics of toxicity, analyzing the real-world events that coincide with significant toxicity peaks and revealing how platform affordances are associated with distinct patterns of toxicity.

\begin{table*}[t!]
\centering

\caption{
Events associated with toxicity peaks in otome game communities from Feb 2024 to Apr 2025.
Each row lists the event type, description, and main target group(s).
Weibo $\Delta$ and Reddit $\Delta$ denote the toxicity change relative to community- and platform-specific averages in~\autoref{figure:toxicity}.
Color codes: \colorbox{di}{\textcolor{black}{green}} for minor fluctuations ($|\Delta|\le 1\%$),
\colorbox{zhong}{\textcolor{black}{yellow}} for moderate increases ($1\%<\Delta\le 3\%$),
and \colorbox{gao}{\textcolor{black}{red}} for notable increases ($\Delta>3\%$).
}
\label{table:time_events}
\tabcolsep 3pt
\renewcommand{\arraystretch}{1.3}
\scalebox{0.8}{
\begin{tabular}{@{}c c p{2.0cm} c c c p{6.7cm} p{4.2cm}@{}}
\toprule
\textbf{No.} & \textbf{Month} & \textbf{Community} & \textbf{Weibo $\Delta$} & \textbf{Reddit $\Delta$} & \textbf{Event Type} & \textbf{Event Description} & \textbf{Main Target Group(s)} \\
\midrule
1 & 2024.02 & \parbox[t]{2cm}{\raggedright General \\ Otome \\ Community} & \cellcolor{gao}+7.17\%  & \cellcolor{zhong}+1.83\%  & In-community conflict & Large-scale debate over ``otome'' standards ignited criticism across multiple communities. \href{https://www.sohu.com/a/752779686_204824}{[Source]} & \parbox[t]{4.6cm}{\raggedright A. players\\C. game developers} \\[2pt]
\hline
2 & 2024.03 & \textit{Mr. Love}         & \cellcolor{gao}+9.31\%  & \cellcolor{zhong}+1.90\%  & In-community conflict & Collab merchandise drew criticism for deceptive design and poor fulfillment. \href{https://www.nfnews.com/content/5yp5qPA43p.html}{[Source]} & \parbox[t]{4.6cm}{\raggedright C. game developers\\F. identity groups} \\[2pt]
\hline
3 & 2024.07 & \textit{Mr. Love}         & \cellcolor{di}-0.61\%   & \cellcolor{gao}+11.94\%   & In-community conflict & Character settings and game developer have caused dissatisfaction among players.& \parbox[t]{4.6cm}{\raggedright C. game developers\\B. NPCs} \\[2pt]
\hline
4 & 2024.08 & \textit{Mr. Love}         & \cellcolor{gao}+27.00\% & \cellcolor{di}-0.09\%    & Game update & Controversial festival card design mass insults emerged. \href{https://36kr.com/p/2904662438386561?utm_source=chatgpt.com}{[Source]} & \parbox[t]{4.6cm}{\raggedright C. game developers} \\[2pt]
\hline
5 & 2024.08 &\parbox[t]{2cm}{\raggedright General \\ Otome \\ Community}  & \cellcolor{gao}+14.89\% & \cellcolor{di}+0.23\%    & External attack & A singer dissed otome players, triggering cross-circle battles involving official game accounts and communities. \href{https://www.thepaper.cn/newsDetail_forward_28651587}{[Source]} & \parbox[t]{4.6cm}{\raggedright C. game developers\\A. players} \\[2pt]
\hline
6 & 2024.10 & \parbox[t]{2cm}{\raggedright General \\ Otome \\ Community}  & \cellcolor{gao}+3.35\%  & \cellcolor{di}-0.54\%    & In-community conflict & Otome game updates spark debates over declining writing quality and fandom culture, raising concerns over lost original intent. \href{https://www.gcores.com/articles/172910}{[Source]} & \parbox[t]{4.6cm}{\raggedright C. game developers\\F. identity groups\\A. players} \\[2pt]
\hline
7 & 2024.12 & \parbox[t]{2cm}{\raggedright General \\ Otome \\ Community}  & \cellcolor{gao}+4.30\%  & \cellcolor{di}+0.16\%    & In-community conflict & Backlash against leading otome games for adopting ``fandom-style'' marketing. \href{https://36kr.com/p/3102211163127555}{[Source]} & \parbox[t]{4.6cm}{\raggedright C. game developers\\F. identity groups\\A. players} \\[2pt]
\hline
8 & 2025.04 & \parbox[t]{2cm}{\raggedright General \\ Otome \\ Community}  & \cellcolor{gao}+3.50\%  & \cellcolor{di}+0.35\%  & Consumer rights protest & On Consumer Rights Day(``315''), widespread protests accused companies and platforms of fraud and deceptive marketing. \href{https://xueqiu.com/1571712278/282698927}{[Source]}\href{http://www.eeo.com.cn/2025/0316/716729.shtml}{[Source]} & \parbox[t]{4.6cm}{\raggedright C. game developers\\E. policymakers\\G. other game-related entities} \\[2pt]
\bottomrule
\end{tabular}
}
\end{table*}

\begin{figure*}[t!]
    \centering
    \includegraphics[
        width=1\linewidth,
        trim=0cm 0.5cm 0cm 0.5cm,
        clip
    ]{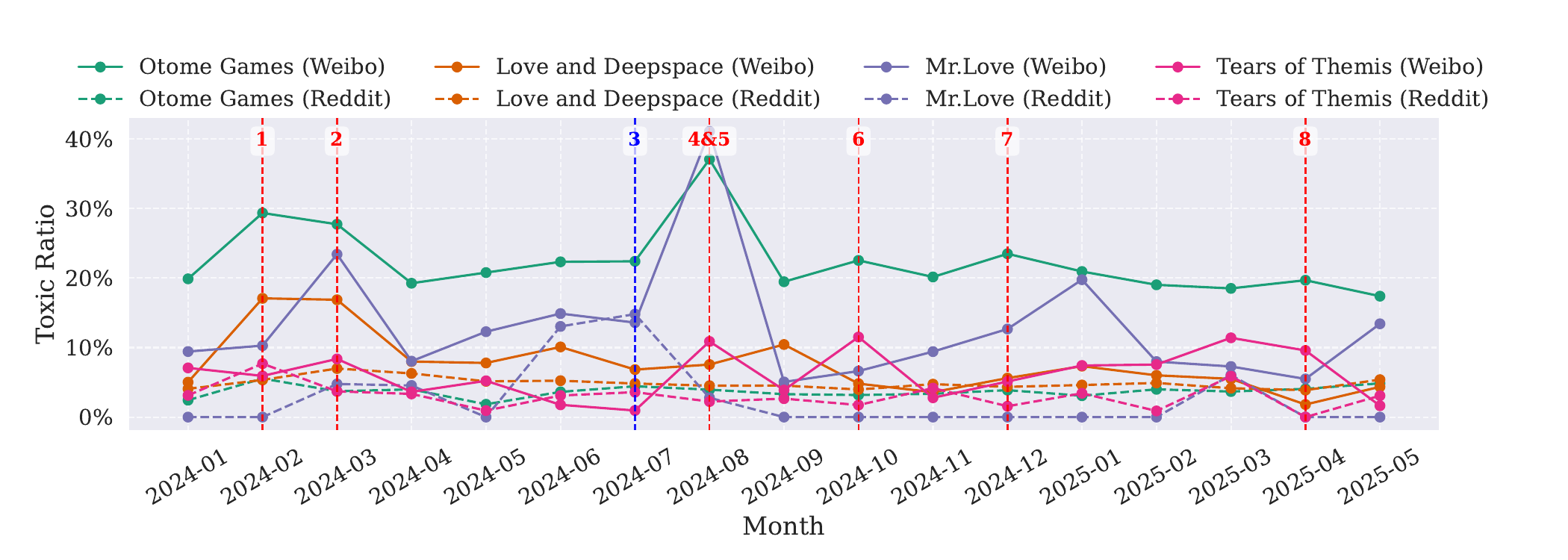}
    \caption{
    Monthly toxicity ratio trends for four otome game communities on Weibo (solid lines) and Reddit (dashed lines). 
    Events are annotated by number and detailed in~\autoref{table:time_events}, with Weibo-related toxicity spikes (Events 1–2, 4–8) shown in red and a Reddit-specific peak (Event 3) in blue.
    }
    \label{figure:toxic_ratio_by_project}
\end{figure*}

\mypara{Methodology}
To investigate the temporal dynamics of toxicity in otome game communities, we conduct a time series analysis of toxicity posts.
Specifically, following previous studies~\cite{JSWSCLBZ24, HA18}, we first normalize the time series of each community by its standard deviation to eliminate fluctuations across communities.
We then apply the peak detection algorithm to identify events associated with toxicity peaks~\cite{BNG11}.
For each identified peak, we gather all posts from the 7 days before and 7 days after it (a 14-day window). 
Posts are grouped into a single event if their top-50 keywords, extracted by the Term Frequency-Inverse Document Frequency (TF-IDF) method~\cite{J04}, have a Jaccard index of at least 0.5.
Our analysis identifies eight events that are significantly associated with toxicity peaks, which are annotated by number in~\autoref{figure:toxic_ratio_by_project} and detailed in~\autoref{table:time_events}.
We manually categorize events into one of four types (i.e., in-community conflict, external attack, game update, consumer rights protest) by validating them through source triangulation using official notices or news reports.
We emphasize the associations reported below are correlational rather than causal.

\mypara{Temporal Analysis}
We find that most events associated with toxicity peaks in otome game communities stem from in-community conflicts.
Among the eight identified events, five fall into this category, focusing on debates over character settings, fan identity, and narrative direction.
In contrast, other event types, such as game updates, external attacks, and consumer rights protests, each account for only one event of increased toxicity.
Take Event \#7 as an example. 
In December 2024, a luxury brand invites NPC characters from an otome game to attend an offline promotional event, and then publicly releases red carpet photos of the NPC characters.
This event coincides with significant backlash from Chinese fans against the game's ``fandom-style'' marketing approach, with the Weibo community showing a notable +4.30\% increase in toxicity.
The initial wave of toxic posts targets the game developers (Target C), criticizing their commercial strategies. 
However, the backlash quickly evolves into a broader ideological debate regarding the cultural norms for women (Target F), specifically, whether female players should be regarded as romantic partners within the game world or as fans.
This abstract discussion is accompanied by infighting among players with divergent perspectives (Target A: players).

Another example is Event \#5, where several rappers publicly disparage otome players on Weibo in August 2024, which coincided with a 14.89\% surge in toxicity. 
The analysis reveals a progressive shift in targets from identity groups to developers and players (see~\autoref{appendix:event5_evolution}).

\begin{tcolorbox}[colback=gray!25!white, size=title,breakable,boxsep=1mm,colframe=white,before={\vskip1mm}, after={\vskip0mm}]
\textbf{Take-Aways:}
Toxicity in otome game communities often surges around real-world events, such as in-community conflicts, game updates, external attacks, and consumer rights protests.
These surges typically follow a pattern: whether associated with internal conflicts or external factors, the toxicity tends to escalate and converge over time.
Within the two platforms studied, toxicity peaks on Weibo and Reddit show little temporal overlap, though this likely reflects the distinct languages and user bases of each platform.
\end{tcolorbox}

\section{Linguistic Features of Otome Toxicity}
\label{section:Linguistic-Strategies}

In this section, we address \textbf{RQ3} by conducting a fine-grained analysis on toxic spans in toxic posts.
We first elaborate on our analysis of toxic spans, and then discuss the specific variant strategies observed in the toxic spans.

\subsection{Toxic Spans}

\begin{CJK}{UTF8}{gbsn}
\begin{table*}[t!]
\centering
\caption{Taxonomy of variation strategies applied to unique toxic spans in otome game communities.
Percentages are computed over unique toxic spans rather than posts.}
\label{table:linguistic_strategies_taxonomy}
\scalebox{0.8}{
\renewcommand{\arraystretch}{1.2}
\begin{tabular}{@{}l c c p{6cm} p{6.5cm}@{}}
\toprule
\textbf{Variation Strategies} & \textbf{\% Weibo} & \textbf{\% Reddit} & \textbf{Description} & \textbf{Examples} \\
\midrule
\textbf{Slang \& Memes} & 26.49\% & 16.17\% & Specialized jargon and memes whose toxic meaning is only clear to community insiders. & ``贱鸟 (insulting nickname for an otome-game company)'', ``国乙之癫 (derogatory label for otome-game players as insane)''/ ``\texttt{salty} (petty, upset)'', ``\texttt{simp} (overly submissive to women/men)'' \\
\addlinespace
\textbf{Letter-Code Abbreviation} & 6.58\% & 5.11\% & Acronyms used to express vulgarity or hostility, often derived from Pinyin or English. & ``sb (shabi, idiot)'', ``tmd (ta ma de, f*ck)'' / ``\texttt{raf} (insulting abbreviation)'', ``\texttt{stfu} (shut the f*ck up)'' \\
\addlinespace
\textbf{Obfuscation via Substitution} & 5.03\% & 0.33\% & Replacing characters with homophones, visually similar characters, or symbols to hide sensitive words. & ``辣鸡 (trash/garbage)'', ``草 (f*ck)'', ``养胃 (euphemism for erectile dysfunction)'' / ``\texttt{shyt} (variant spelling of shit)'' \\
\addlinespace
\textbf{Emoji Substitution} & 1.79\% & 1.16\% & Using emojis to convey negative sentiment, sarcasm, or to stand in for offensive words. & \emoji{horse} (homophone for mother in insults), \emoji{poop} (calling someone ``shit'') / \emoji{vomiting} (expressing disgust) \\
\midrule
\textbf{Any Variant Strategy Applied} & \textbf{39.89\%} & \textbf{22.77\%} & & \\
\bottomrule
\end{tabular}
}
\end{table*}
\end{CJK}

\mypara{Methodology}
As mentioned in~\autoref{section:evaluation}, when the LLM-driven detectors determine that a given sample is toxic, it also outputs a field named \texttt{toxicity\_reason}, which identifies the specific toxic spans contributing to the toxic classification.
To leverage this information, we extract the explanatory text from the \texttt{toxicity\_reason} field for all toxic posts.
We then tokenize both explanatory text and original post using platform-appropriate methods: for Chinese posts from Weibo, we use Jieba~\cite{jieba}, for English posts from Reddit, we use a regular expression to extract alphabetic tokens of two or more letters. 
Finally, we compute the intersection of the two corresponding token sets for each post.
This approach allows us to retain only the tokens from the original post that the model deems toxic, thereby grounding our analysis in the actual content of the text.
In the end, we identify 4,013 unique Chinese and 606 unique English toxic spans. 

\begin{figure}[t!]
    \centering
    \includegraphics[width=1\linewidth]{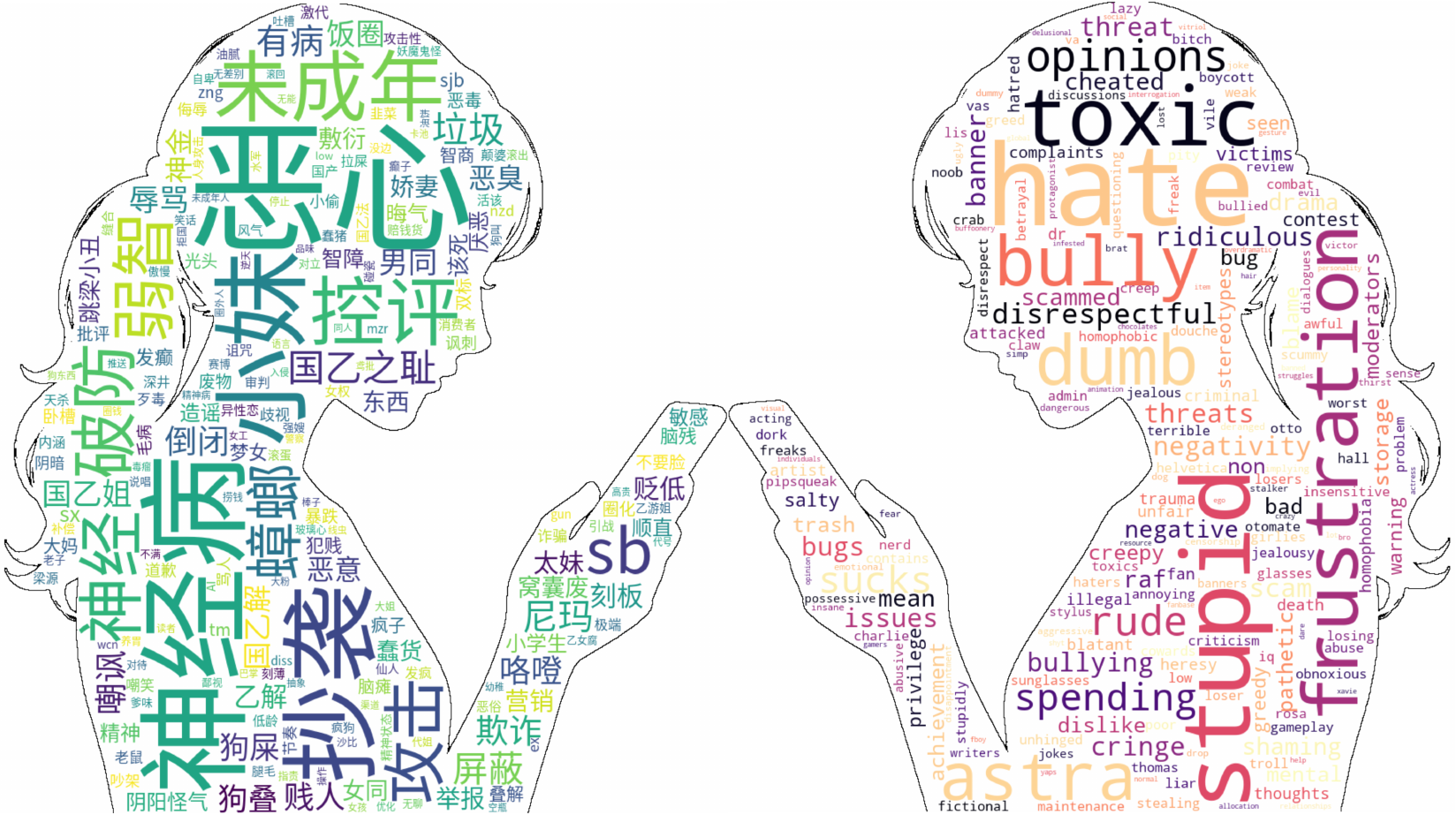}
    \caption{Word cloud of toxic spans in the otome game community on Weibo (left) and Reddit (right).}
    \label{figure:wordcloud}
\end{figure}

\mypara{Visual Exploration}
We then visualize the identified toxic spans through word clouds, as shown in~\autoref{figure:wordcloud}. 
We find that common slurs generally appear on both platforms, such as ``\begin{CJK}{UTF8}{gbsn}弱智\end{CJK} (stupid, dumb),'' ``\begin{CJK}{UTF8}{gbsn}恶心\end{CJK} (cringe),'' and ``\begin{CJK}{UTF8}{gbsn}破防\end{CJK} (frustration).''
However, Weibo also includes jargon associated with its ``fan circle'' culture,\footnote{Fan circle culture (``\begin{CJK}{UTF8}{gbsn}饭圈\end{CJK}'' in Chinese) refers to the organized networks of fans who actively promote and support celebrities or idols on social media, often through coordinated activities like comment control, voting, and sometimes toxic behaviors~\cite{MWC23}.} such as ``\begin{CJK}{UTF8}{gbsn}饭圈\end{CJK} (fan circle)'' and ``\begin{CJK}{UTF8}{gbsn}控评\end{CJK} (comment control).''
It also contains derogatory terms targeting specific fan groups in the otome game community, such as ``\begin{CJK}{UTF8}{gbsn}小妹\end{CJK} (derogatory words refer to young or immature female fans, implying they are childish and overly aggressive in online disputes)'' or ``\begin{CJK}{UTF8}{gbsn}国乙姐\end{CJK} (a pejorative term for fans of domestic Chinese otome games, suggesting they have low standards, accept subpar content, or exhibit biased and combative behavior in the community),'' as well as pejorative nicknames for game companies like ``\begin{CJK}{UTF8}{gbsn}狗叠\end{CJK} (Dog-Paper Games, a mocking alteration of Paper Games'' where ``dog'' implies the company is greedy, neglectful, or produces low-quality work that frustrates players).''
In contrast, Reddit contains terms expressing dissatisfaction with the game's content or the company's operations, including ``spending,'' ``scam,'' ``bugs,'' and ``greedy.''
This suggests that toxic vocabulary differs across platforms, which may reflect differences in user bases and cultural contexts.

\subsection{Variation Strategies in Toxic Spans}
When checking the toxic spans, we observe multiple variation strategies, such as homophonic substitutions and abbreviations.
These strategies, as suggested by previous literature~\cite{XHCL24, RMAY25}, may make toxic expressions harder for lexical moderation systems to detect and may also align with platform-specific language norms.
Therefore, to better understand these practices, we perform an iterative coding on the toxic spans.
In the end, we identify four variation strategies, which are slang\&memes, letter-code abbreviation, obfuscation via substitution, and emoji substitution, summarized in~\autoref{table:linguistic_strategies_taxonomy}.
We find that users on both platforms most commonly use variation strategies like slang\&memes and letter-code abbreviations.
On Weibo, 26.49\% of toxic spans use slang\&memes, while 6.58\% leverage letter-code abbreviations. 
Similarly, on Reddit, 16.17\% of toxic spans contains slang\&memes, and 5.11\% employ letter-code abbreviations.
At the unique-span level, the use of at least one variation strategy is significantly more prevalent on Weibo, appearing in 39.89\% of unique toxic spans, compared to 22.77\% on Reddit.
Notably, these estimates should be interpreted as lower bounds. 
Our analysis begins with posts identified as toxic by the selected
classifiers. 
Therefore, toxic posts that successfully evade detection are absent from the analyzed set and may contain additional or more sophisticated variation strategies. 
The reported percentages characterize model-detected unique toxic spans rather than the complete population of toxic content.
This span-level result suggests that toxic expressions on Weibo more often rely on linguistic variation, potentially as a means of complicating moderation or signaling in-group identity.

\begin{tcolorbox}[colback=gray!25!white, size=title,breakable,boxsep=1mm,colframe=white,before={\vskip1mm}, after={\vskip0mm}]
\textbf{Take-Aways:}
Toxic posts in otome game communities show distinct linguistic patterns across platforms. 
On Weibo, nearly 40\% of unique toxic spans use variation strategies like slang, abbreviation, and substitution.
In contrast, Reddit users tend to express toxicity more directly, relying less on variation strategies. 
These differences are consistent with platform-specific patterns in how toxicity is expressed, not only in what it concerns.
\end{tcolorbox}

\section{From Individual Toxicity to Potential Coordination}
\label{section:case_study_weibo}

Building on our analysis of individual toxic posts, we investigate whether toxic posts form clusters exhibiting potential coordination in otome game communities, thereby answering \textbf{RQ4}.

\mypara{Potential-Coordination Detection}
Prior work defines coordinated harassment as collective abuse in which multiple actors collaboratively post similar toxic content against a target~\cite{TABBBCDDKKMMRS21}.
However, public behavioral traces cannot directly represent participants' intent, so our method identifies only potential coordination.
Following previous studies~\cite{PHTTFM21}, we first tokenize each post and transform it into a TF-IDF vector.
Then, we compute the pairwise cosine similarity between all post vectors. 
We group posts into a single cluster if their cosine similarity score is 0.80 or higher. 
This threshold follows prior similarity-based studies and balances the identification of semantically similar content with allowance for minor textual variation~\cite{AF19}.
This process identifies 1,375 clusters.
We then retain only those clusters containing at least one post classified as toxic by the detector selected in~\autoref{section:evaluation}, focusing our analysis on potentially harmful clustered behavior.
This filtering yields 191 candidate clusters exhibiting potential coordination.
We manually review all 191 candidate clusters based on textual-template reuse, shared hashtags, target consistency, and temporal concentration.
In this review, 49.74\% of the candidate clusters exhibit template-like or near-duplicate wording, and 28 accounts contribute to at least two clusters.
One cluster appears to reflect only independent but similar reactions.
We nevertheless retain it in the candidate set, as our procedure reports flagged candidates rather than confirmed coordination.

\begin{figure}[t]
  \centering
  \begin{subfigure}[t]{0.328\linewidth}
    \centering
    \includegraphics[width=\linewidth]{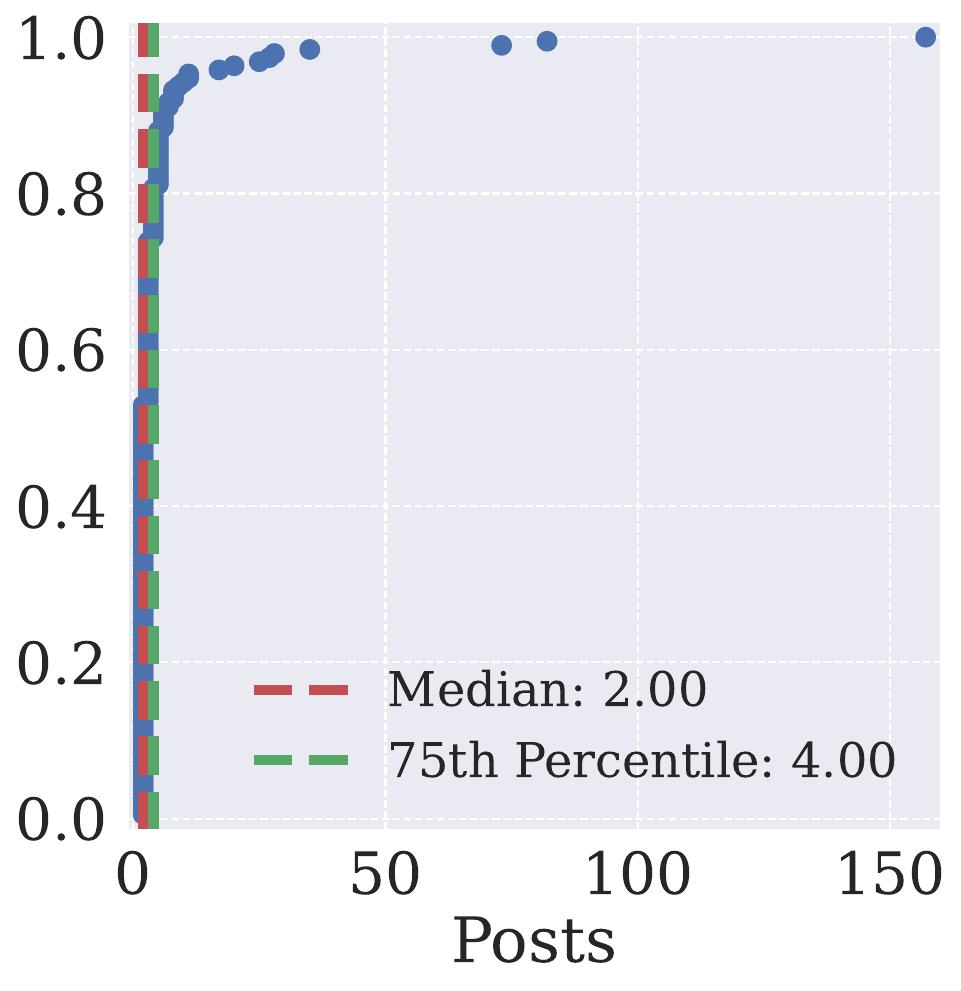}
    \caption{Posts}\label{figure:cdf_posts}
  \end{subfigure}\hfill
  \begin{subfigure}[t]{0.328\linewidth}
    \centering
    \includegraphics[width=\linewidth]{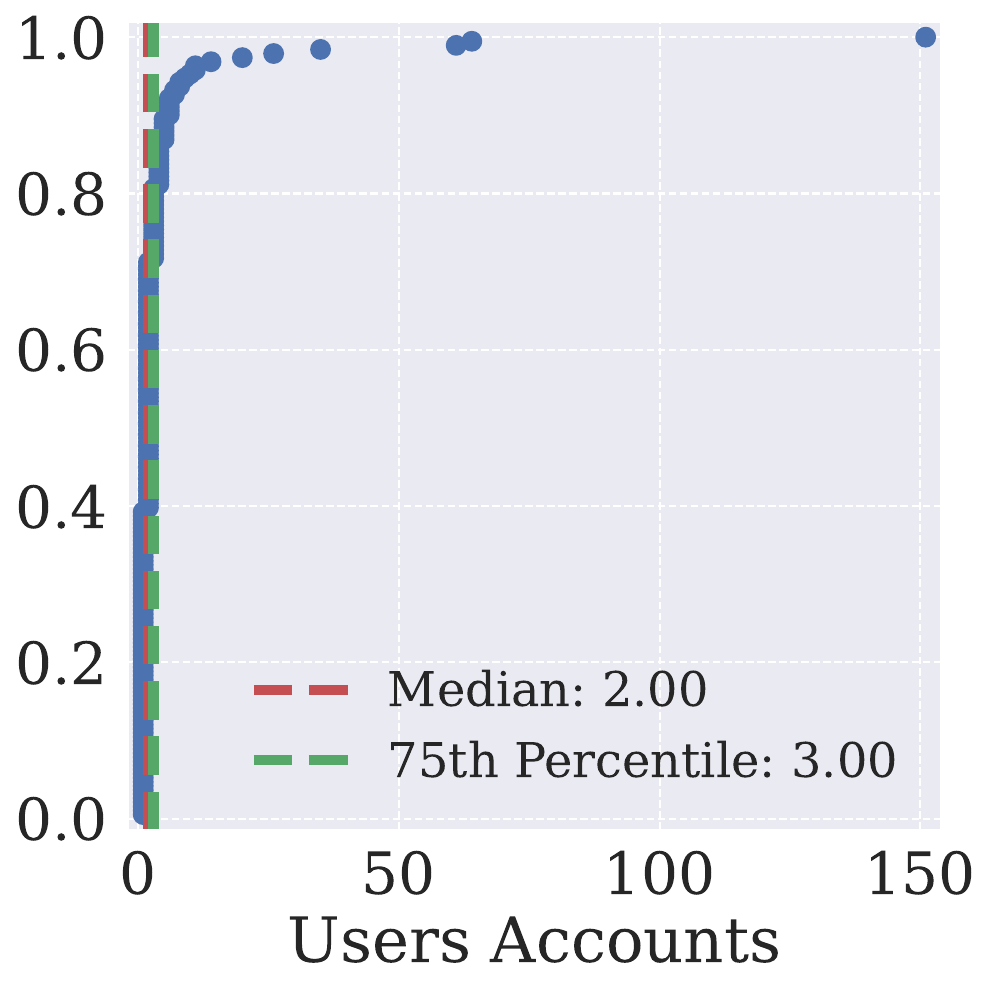}
    \caption{User Accounts}\label{figure:cdf_users}
  \end{subfigure}\hfill
  \begin{subfigure}[t]{0.328\linewidth}
    \centering
    \includegraphics[width=\linewidth]{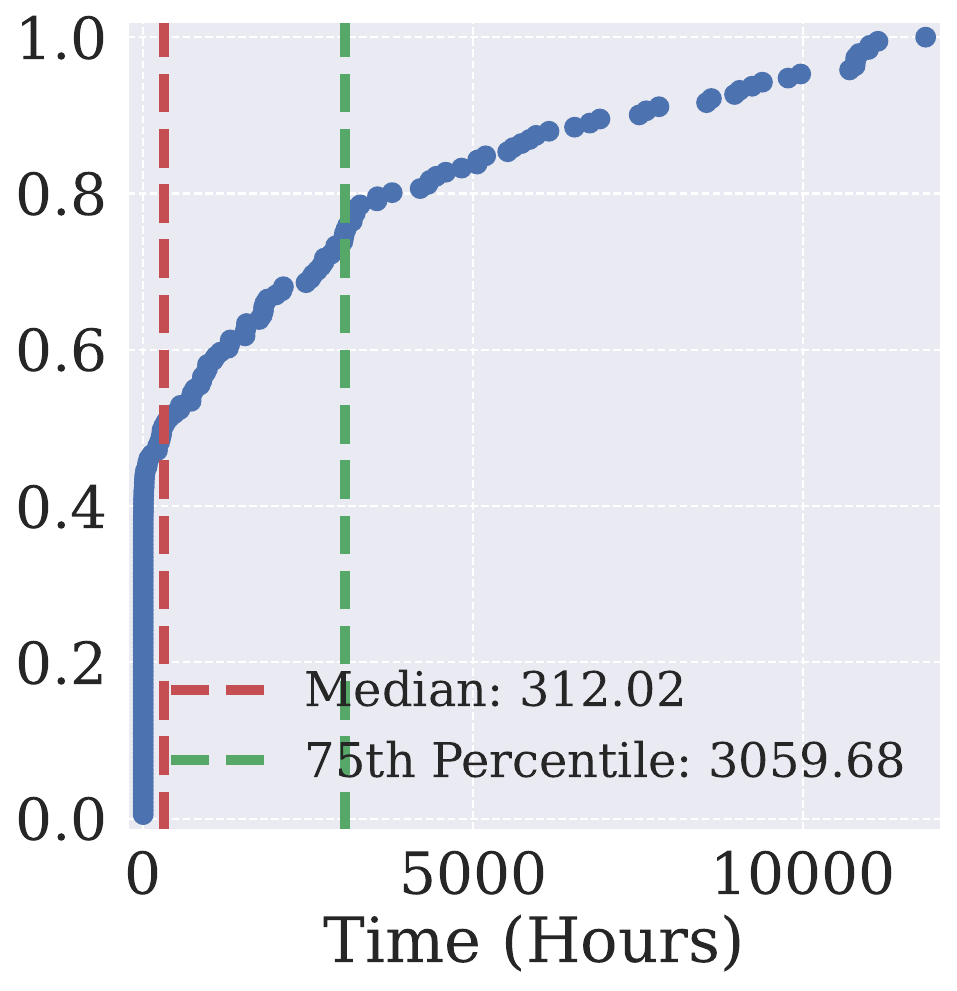}
    \caption{Time Span}\label{figure:cdf_time}
  \end{subfigure}
\caption{CDF plots of potential-coordination clusters.}
  \label{figure:cdf_plots}
\end{figure}

\mypara{Potential-Coordination Cluster Analysis}
\autoref{figure:cdf_plots} shows the CDFs of the
potential-coordination clusters across three dimensions: the number of posts, the number of contributing accounts, and the time span. 
Most clusters contain few posts, with a median of 2 and 75\% containing no more than 4 posts. 
The number of contributing accounts has a median of 2 and a 75th percentile of 3. 
The median cluster duration is approximately 13 days, while the 75th percentile is close to 128 days.
Thus, most clusters are small, while a minority persist for substantially longer periods.

\mypara{Categorization of Potential-Coordination Clusters}
We categorize potential-coordination clusters along two descriptive dimensions: \textit{scale}, measured by the number of posts and contributing accounts, and \textit{persistence}, measured by the cluster time span. 
A cluster is categorized as High Scale if it exceeds the 75th percentile in either posts or contributing accounts.
Otherwise, it is categorized as Low Scale. 
Clusters lasting longer than the median duration of 312 hours are categorized as Long Persistence, while the remaining clusters are categorized as Short Persistence. 
Combining these dimensions produces four descriptive categories, as shown in~\autoref{figure:attack_taxonomy}.

\autoref{table:attack_strategy_crosstab} in~\autoref{appendix:event5_evolution} shows the distribution of the four scale persistence categories across target groups.
The most common category is low scale and short persistence (42.93\%), primarily targeting game developers (79.27\%).
The high-scale, short-persistence category is the least common (6.81\%) but has the highest proportion targeting game developers (92.31\%).
In contrast, long-persistence clusters exhibit a more diverse target distribution.
Among low-scale, long-persistence clusters (32.46\%), game developers remain the most frequent target (48.39\%), followed by unknown targets (19.35\%) and players (17.74\%).
High-scale, long-persistence clusters (17.80\%) also most frequently target game developers (47.06\%), followed by unknown targets (32.35\%), players (8.82\%), and NPCs (8.82\%).
Overall, game developers are the most frequent target across all four categories, while long-persistence clusters involve a broader range of targets.

\begin{figure}[t!]
    \centering
    \includegraphics[width=0.5\linewidth]{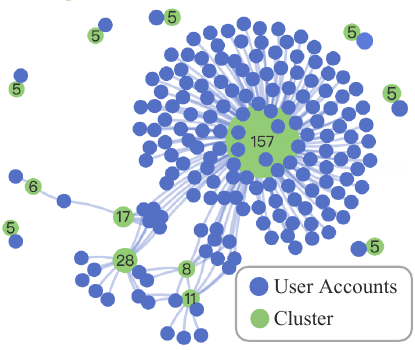} 
    \caption{
Bipartite network of potential-coordination clusters and the accounts that contributed to them.
An edge indicates that an account contributed at least one post to a cluster.
The number represents the total number of posts in the cluster.
    }
    \label{figure:core_threat_network}
\end{figure}

\mypara{Case Study: Repeated Participation Across Clusters}
Because 92.31\% of the high-scale, short-persistence clusters target game developers, we construct a bipartite network connecting these clusters with the accounts that contributed posts to them, as shown in~\autoref{figure:core_threat_network}.
While some clusters appear isolated, a persistent group of user accounts is repeatedly involved in six clusters.
This repeated participation is an observable pattern.
It does not establish a coordinated effort or exclude independent participation.
We then manually reviewed the user accounts involved in these clusters.
Interestingly, we find that one of the most central nodes (user accounts) that connects multiple clusters is a verified community influencer with over 2,000 followers, and several other highly active accounts are created shortly before the relevant posting periods and exhibit high activity only during those periods.
These accounts post highly similar or near-identical content, which is consistent with template reuse or shared messaging. 
We distinguish these accounts from passionate fans who also post frequently.
A sensitivity analysis across cosine similarity thresholds is reported in~\refappendix{appendix:campaign_sensitivity}.

\begin{tcolorbox}[colback=gray!25!white, size=title,breakable,boxsep=1mm,colframe=white,before={\vskip1mm}, after={\vskip0mm}]
\textbf{Take-Aways:}
Our similarity-based procedure flags 191 potential-coordination clusters in otome game communities.
Most clusters are low scale, and they are roughly evenly split between shorter and longer persistence.
Game developers are the most frequent target across categories, while long-persistence clusters span a broader range of targets.
Several user accounts appear repeatedly across multiple clusters, including a verified influencer with over 2,000 followers and several accounts created shortly before the relevant posting periods.
These are observable patterns that warrant further investigation.
Our data do not establish coordination or participants' intent.
\end{tcolorbox}

\section{Discussion}
\label{section:discussion}

Our analysis, moving from large-scale statistical measurements to an examination of potential-coordination clusters, shows that toxicity in otome game communities is not a monolithic failure of civility.
The contrast between Weibo and Reddit is consistent with qualitatively different regimes of conflict.

\mypara{Cross-Platform Patterns of Conflict}
On Weibo, 55.4\% of the coded toxic posts involve factional attacks or identity-marking language, and 42.4\% use in-group slang.
This pattern is consistent with identity signaling within factionalized fan subcultures, but the post content alone does not establish users' motivations.
Insults frequently target rival groups, while ``Super Topics'' may amplify these conflicts.
Because our design does not include a controlled comparison between Super Topic and non-Super-Topic posts, their possible amplifying role remains a hypothesis for future research.
On Reddit, 51.0\% of the coded toxic posts concern parasocial or consumer/developer grievances.
Users frequently express grievances about writing or monetization, and toxic speech may serve as leverage in these negotiations.
The platform's persistent, threaded forum structure may support this process by enabling sustained argumentation and accumulation of dissent, although our analysis does not isolate this platform effect from language, user composition, or moderation.

\mypara{Implications for Socio-Technical Security}  
These differences suggest that moderation cannot rely on uniform strategies.
For Weibo, the prevalence of factional attacks and identity-marking language motivates evaluating reversible interventions such as algorithmic de-amplification or reputation systems.
For Reddit, future evaluations could test whether credible grievance channels and tools that separate strong criticism from identity-based harassment reduce developer-targeted toxicity, since heavy censorship may reinforce distrust.
We develop concrete experimental designs for both strategies in \refappendix{appendix:gov:experiments}, and propose evasion-aware detection modules (\refappendix{appendix:gov:evasion}) and early-warning indicators (\refappendix{appendix:gov:earlywarning}) informed by our empirical findings.

\mypara{Limitation and Future Work}
Our analysis is confined to Weibo and Reddit, and future work could explore other platforms such as Twitter or Discord to provide a more holistic view of the otome game community ecosystem. 
Then, our data spans from January 2024 to May 2025, a longitudinal study over a longer timeframe could reveal evolving toxicity patterns and community norms. 
Methodologically, while our LLM-driven detectors perform well, they are not perfect, and future research could focus on developing models more robust to the creative and evasive linguistic strategies we identify. 
Moreover, our study focuses only on Chinese and English, which are the most prominent languages with the largest player bases.
As the first study of otome game communities, we chose to focus on them, while leaving the investigation of other languages to future work.

\section{Conclusion}
We conduct the first large-scale investigation of toxicity in otome game communities, a space that remains underexplored despite its growing global prominence. 
By introducing \OurMethod, we systematically analyze the toxicity in otome game communities across multiple dimensions, including its prevalence, targeted groups, interaction patterns, temporal dynamics, linguistic characteristics, and observable patterns of potential coordination.
We hope this work will advance further research into the sociotechnical factors underlying online toxicity and contribute to the development of safer, more inclusive digital environments.

\bibliographystyle{plain}
\bibliography{normal_generated_py3}

\appendix 

\section{Open Science}
\label{appendix:open_science}

We publicly release the manually annotated dataset on Hugging Face at \AnnotatedDatasetURL.
The repository contributions on Hugging Face contain CSV files comprising 4,308 manually annotated posts (2,255 from Weibo, 2,053 from Reddit) across six communities per platform, with toxicity labels and target labels.

\section{Ethical Considerations}
\label{appendix:ethics}

Our research exclusively relies on publicly available data from the social media platforms Weibo and Reddit.
Nevertheless, public availability does not eliminate users' contextual privacy expectations, particularly in niche fan communities such as those studied here, where discussions are highly specialized and members may not anticipate research use~\cite{FZPGJ24, FP18, PJGFZ21}.
We do not collect any private user information, nor do we attempt to circumvent any platform's privacy controls.
We did not contact community moderators prior to data collection.
Our study collected only publicly accessible posts and did not interact with users or intervene in community discussions.
Nevertheless, we acknowledge that the absence of such consultation limits our ability to account for community-specific expectations regarding research use.
To protect the privacy of individuals whose posts appear in our dataset, we remove or pseudonymize all personally identifiable information, such as usernames and user IDs, during the preprocessing stage.
Posts marked as deleted or removed were excluded during data cleaning on both platforms.
We do not quote such posts or include them in released datasets or example sets.
To reduce re-identification risk, all examples shown in this paper are rewritten by the authors and do not expose individual posts.
Throughout, we omit usernames, user IDs, URLs, and post titles.
These measures follow the practices of previous studies~\cite{FZPGJ24}.
However, as shown in prior work~\cite{R22}, they reduce but cannot completely eliminate the possibility of re-identification.

Two of the paper's authors, who possess deep domain expertise as long-term otome game players, performed the manual annotation of toxic posts.
By conducting the annotation internally, we ensure that no external annotators are exposed to potentially harmful or distressing content. 
The authors are fully aware of the nature of the content and consent to the task, with the understanding that they may take breaks or cease annotation at any time to mitigate potential psychological impact. 
At our institution, ERB review was not required for this study, which analyzes only publicly available data and does not involve direct interaction with or intervention in the lives of individuals. 
We nevertheless followed the ethical considerations outlined by our institution's ethics framework, including stakeholder impacts, personal data handling, and dual-use risks.

We recognize the dual-use potential of our findings, particularly the analysis of linguistic evasion strategies in~\autoref{section:Linguistic-Strategies} and potential-coordination clusters in~\autoref{section:case_study_weibo}. 
Although these insights could be misused to improve evasion or coordinated abuse, we mitigate this risk by withholding the full corpus and limiting released materials to those necessary for evaluating the study.
Our goal is to foster the development of more robust, context-aware moderation systems. 
To this end, we commit to responsibly disclosing our findings to the safety and security teams at Weibo and Reddit.

\section{Sample Size Determination for Annotation}
\label{appendix:sampling_stats}

To obtain statistically reliable manual labels, we derive the minimum sample size required to estimate a population proportion with a 95\% confidence level and a margin of error $E=0.05$. 
We treat toxicity as a Bernoulli variable with maximal variance ($p=0.5$) and apply the standard formula with $Z = 1.96$, $n = \frac{Z^{2} \, p(1-p)}{E^{2}}$.
This calculation yields an initial sample size of $n=384$ for an infinitely large population.
Because each community in our corpus contains a finite number of posts, we apply the finite-population correction (FPC) to adjust the sample size to $n_{\text{adj}} = \frac{n}{1 + \frac{n-1}{N}}$.
For example, for the \textit{Mr. Love} super topic with $N = 13{,}563$ posts and an initial sample size of $n = 384$, the adjusted size is
$n_{\text{adj}} \approx \frac{384}{1 + \frac{384-1}{13{,}563}} \approx 374.$

\section{Keyword Search Strategy}
\label{appendix:keyword_filter}

To maximise recall while keeping noise low, we adopt a keyword search strategy on Weibo.
Reddit data are directly obtained via subreddit rather than keyword search and are therefore excluded from this discussion.

\mypara{Keyword List Construction}
To ensure our data collection on Weibo was both comprehensive and relevant, we develop the keyword list through an iterative process. 
This process is guided by the domain expertise of the authors, who are long-term otome game players.
We begin with a foundational set of keywords, including the official game titles and generic terms like ``otome game'' (\begin{CJK}{UTF8}{gbsn}乙女游戏\end{CJK}).
We then systematically expand this list to include common abbreviations and variations frequently used within the player communities, such as \begin{CJK}{UTF8}{gbsn}乙游\end{CJK} (abbr.~``otome games'') and \begin{CJK}{UTF8}{gbsn}国乙\end{CJK} (``domestic otome''). 
Specifically, we searched for posts containing core otome-related terms such as \begin{CJK}{UTF8}{gbsn}乙女游戏\end{CJK} (``otome games''), its abbreviation \begin{CJK}{UTF8}{gbsn}乙游\end{CJK}, \begin{CJK}{UTF8}{gbsn}国乙\end{CJK} (``domestic otome''), \begin{CJK}{UTF8}{gbsn}乙女\end{CJK}, and \begin{CJK}{UTF8}{gbsn}乙女向\end{CJK} (``otome-oriented''), as well as their English counterparts ``otome'' and ``otome games''. 
For comparison, control queries used more general gaming terms such as \begin{CJK}{UTF8}{gbsn}游戏\end{CJK} (``games''), ``game'', and ``games''. 
We excluded additional modifiers like \begin{CJK}{UTF8}{gbsn}手游\end{CJK} (``mobile game'') or genre tags (e.g., ``RPG'') to reduce noise.

\mypara{Query Execution}
We execute all keyword queries via the \texttt{weibo-search} tool.
We retain only posts originating from \emph{Super Topics} (\begin{CJK}{UTF8}{gbsn}超话\end{CJK}) with games \textit{Love and Deepspace}, \textit{Mr. Love: Queen's Choice}, \textit{Tears of Themis}, and \textit{Genshin Impact}, and for all communities discard retweets and administrative notices, and de-duplicate based on post \texttt{id}. 

\mypara{Coverage Check}
To validate the keyword list, we manually reviewed a random 1\% sample of the collected posts. 
The review shows that genre-specific precision exceeds 99\%, confirming that the keyword list was sufficient to capture the relevant discourse.

\begin{CJK}{UTF8}{gbsn} 
\begin{table*}[t!]
\centering
\caption{Examples of toxic and non-toxic posts on Weibo and Reddit. Examples are rewritten by the authors.}
\label{table:toxicity_examples}
\tabcolsep 3pt
\scalebox{0.6}{
\begin{tabularx}{1.6\textwidth}{p{.06\linewidth}l|X p{.18\linewidth} X|X p{.18\linewidth} X}
\toprule
& \textbf{Label} & \textbf{Weibo Example} tox=0 & \textbf{Weibo Example Translation} tox=0 & \textbf{Reddit Example} tox=0 & \textbf{Weibo Example} tox=1 & \textbf{Weibo Example Translation} tox=1 & \textbf{Reddit Example} tox=1 \\
\midrule
\multirow{1}{*}{\textbf{Toxicity}} &
0/1 &
感觉国乙男主的人设越来越同质化了 &
The male leads in domestic otome games feel more and more homogeneous. &
\texttt{Can someone explain how the affinity system works? I just started.}  &
玩这种游戏的人脑子都有问题吧 &
Anyone who plays this kind of game must have something wrong with their head. &
\texttt{Everyone in this fandom is a pathetic loser, just shut up already.} \\
\midrule
\multirow{8}{*}{\textbf{\shortstack{Target\\Groups}}} &
A. players &
有姐妹玩过这个活动吗，积分怎么攒比较快 &
Has anyone here played this event? What's the fastest way to farm points? &
\texttt{Can any experienced players share tips for the timed stages?} &
有些乙游玩家真是nt，走到哪儿都要吵一架 &
Some otome players are just \texttt{nt} (\textit{naotan}, brain-dead), starting a fight wherever they go. &
\texttt{Otome fans are clueless idiots who turn every thread into a fight.} \\
\cmidrule{2-8}
& B. NPCs &
这个男主新剧情写得挺温柔的，我还挺喜欢 &
The male lead comes across as really gentle in the new chapter, and I quite like it. &
\texttt{I liked this character in the new chapter, though some of his decisions frustrated me.} &
这个男主真的恶心，看见他就烦 &
This male lead is genuinely disgusting. I'm sick of the sight of him. &
\texttt{That love interest is a disgusting loser, and I'm sick of seeing him.}\\
\cmidrule{2-8}
& \shortstack{C. game\\developers} &
这次更新内容太少，礼包定价也不太合理 &
There's too little content in this update, and the bundle pricing isn't reasonable either. &
\texttt{This update offers too little content for its price.} &
策划就是一群只会圈钱的辣鸡 &
The planners are nothing but a pack of money-grubbing \textit{laji} (homophone of ``garbage''). &
\texttt{The devs are greedy morons who treat players like wallets.} \\
\cmidrule{2-8}
& \shortstack{D. platform\\moderators} &
我刚发的帖子为什么没通过审核 &
Why didn't the post I just submitted pass review? &
\texttt{Can a mod explain why my post got removed?} &
超话管理有病吧，这都要管 &
Is the Super Topic admin sick in the head? They have to police even this. &
\texttt{The mods here are useless bullies who just abuse their authority.} \\
\cmidrule{2-8}
& E. policymakers &
想问下现在的分级标准是怎么界定的 &
Could I ask how the current rating standards are actually defined? &
\texttt{Did the regional regulators explain why this title isn't available here?} &
定这种规则的都是些不懂还瞎管的傻福 &
The ones setting rules like this are meddling \textit{shafu} (character substitution for a common insult) who don't understand a thing. &
\texttt{The regulators who wrote these rules are brainless fools.}  \\
\cmidrule{2-8}
& \shortstack{F. identity\\groups} &
不同性别的玩家可能会有不同的角色偏好 &
Players of different genders may have different character preferences. &
\texttt{Players of different genders may connect with the story differently.} &
女的根本不配谈游戏设计，都闭嘴吧 &
Women aren't fit to talk about game design at all. Just shut up, all of you. &
\texttt{Women are too stupid to understand game design and should stay quiet.} \\
\cmidrule{2-8}

& \shortstack{G. other\\game-related\\entities} &
感觉这位配音的声音不太适合这个角色 &
I feel this voice actor's voice doesn't quite suit the character. &
\texttt{Honestly, I don't think this voice actor's performance fits the character.} &
给这种角色配音的真辣眼睛，赶紧滚 &
Whoever voices a character like this is an eyesore. Get lost. &
\texttt{That voice actor is talentless trash and should just quit.}   \\
\cmidrule{2-8}

& H. unknown &
这次确实有点让人失望 &
This really was a bit of a letdown. &
\texttt{Well, that was disappointing.} &
脑子有泡吧，服了 &
Must be out of their mind. Unbelievable. &
\texttt{What an insufferable clown. Absolutely useless.}\\
\bottomrule
\end{tabularx}
}
\end{table*}
\end{CJK}

\section{Boundary Cases for Toxicity Annotation}
\label{appendix:toxicity_boundary}

To avoid conflating harmful toxicity with mere disagreement, we apply explicit boundary rules during annotation.
We distinguish strong criticism from toxicity by the target and form of the expression.
Profanity used to criticize a product or narrative is not sufficient for a toxicity label, whereas direct insults, slurs, threats, or harassment toward people, groups, or characters are considered toxic.
We discuss these boundary rules during the pilot study and incorporate them into the final codebook.
Representative boundary cases are shown in~\autoref{table:toxicity_boundary_cases}.

Two boundary types account for most disagreements during the pilot study.
The first is strong criticism of an artifact.
Profanity and intensity do not by themselves shift the target from the artifact to a person: we label a post toxic only when a demeaning term is predicated of a person, group, or character.
Under this rule, ``this banner is garbage'' remains non-toxic while ``the people who made this banner are garbage'' is toxic, even though both express the same dissatisfaction.
A related case is criticism that reproduces a slur: quoting or reporting a slur in order to object to it is not labeled toxic, whereas using the same term to characterize a person or group is.
The second is language directed at fictional characters.
We separate evaluation of narrative craft, which we do not label as toxic, from demeaning language predicated of a character, which we do.
\autoref{section:RQ1} discusses why the latter is relevant to community interaction even though characters cannot themselves be harmed.
We acknowledge that these boundaries are not always crisp and that a small number of posts remain genuinely ambiguous.

\begin{table}[t!]
\centering
\caption{Boundary cases used to distinguish disagreement from toxicity. Examples are rewritten by the authors.}
\label{table:toxicity_boundary_cases}
\scalebox{0.8}{
\begin{tabular}{@{}p{2.2cm}p{3.0cm}p{1.2cm}p{2.8cm}@{}}
\toprule
\textbf{Case} & \textbf{Example} & \textbf{Label} & \textbf{Rationale} \\
\midrule
Negative opinion & \texttt{I dislike this storyline.} & Non-toxic & Expresses preference without attacking a target. \\
Consumer complaint & \texttt{This event is overpriced.} & Non-toxic & Criticizes monetization without derogatory language. \\
Aggressive complaint & \texttt{The developers are greedy clowns.} & Toxic & Uses demeaning language toward developers. \\
Product criticism with profanity & \texttt{This banner design is fucking terrible.} & Non-toxic & Profanity intensifies criticism of an artifact, and no demeaning term is applied to a person or group. \\
Criticism naming a slur & \texttt{Why name an item after a word used to demean women?} & Non-toxic & Reports a slur in order to object to it rather than directing it at anyone. \\
Character-directed critique & \texttt{This character's arc was written lazily.} & Non-toxic & Evaluates narrative craft without demeaning language. \\
Character-directed abuse & \texttt{This character is a disgusting creep.} & Toxic & Predicates a demeaning term of a character rather than evaluating the writing. \\
Group attack & \texttt{Otome players are brain-dead.} & Toxic & Attacks a player community with derogatory language. \\
Counter-speech & \texttt{Stop insulting otome players.} & Non-toxic & Condemns harassment rather than attacking a target. \\
Ambiguous profanity & \texttt{What the hell is this banner?} & Non-toxic & Contains profanity but no direct demeaning target. \\
\bottomrule
\end{tabular}
}
\end{table}

\section{Prompt Design Details}
\label{appendix:prompt}

We design all prompts to classify user-generated content with respect to toxicity and target groups.

\begin{itemize}
  \item \textbf{Prompt 1: CoT (Chain-of-Thought).}  
  We adapt a multi-step prompt from \textit{HateCoT}~\cite{NI24}, which decomposes the task into four steps: identifying mentioned targets, checking toxicity, matching attack direction, and generating a compact label.
This design makes intermediate reasoning externally visible.

  \item \textbf{Prompt 2: Concise Definition.}  
  We define toxicity and target categories directly and instruct the model to output two fields: \texttt{``toxicity''} and \texttt{``target''}, optionally with a brief explanation.  
  We provide two versions: Chinese examples drawn from Weibo posts, and English examples adapted for Reddit input.

  \item \textbf{Prompt 3: Reasoning.}  
  We extend Prompt 2 by requiring a short reasoning for both labels.
This design introduces two optional fields, \texttt{``toxicity\_reason''} and \texttt{``target\_reason''}, alongside the two classification outputs.
As in Prompt 2, we provide both Chinese and English example variants.
\end{itemize}

The only structural difference between Prompt 2 and Prompt 3 lies in the addition of reasoning fields, which encourages the model to justify its predictions.

The in-context examples shown below are rewritten by the authors for presentation.
The prompts used in the experiments contain posts drawn from the collected corpus but outside the annotated evaluation set.
The exact prompt text is included in the research artifacts described in Appendix~\ref{appendix:open_science}.

\begin{tcolorbox}[title=Prompt 1: CoT, colback=white!97!gray, boxrule=0.5pt,  breakable]
\begin{Verbatim}[fontsize=\scriptsize,breaklines=true,breaksymbolleft={},breaksymbolright={}]
As a specialist annotator focused on language analysis in otome game communities, your role is to carefully examine user posts from platforms (e.g., Weibo, Reddit). Your expertise is essential in identifying potentially toxic posts
Step 1: Identify Mentioned Targets (Q1)
Please determine which of the following categories the comment refers to:
Target Categories (A–H):
   - `A`: players - `B`: NPCs - `C`: game developer
   - `D`: platform moderators  - `E`: policymakers
   - `F`: identity groups - `G`: other game-related entities 
   - `H`: unknown
Step 2: Check for Toxicity (Q2)
Does the comment contain any of the following? 
- `0` — No. - `1` — Yes.
Step 3: Directional Attack Matching (Q3)
If Q2 = 1 (Yes), determine which specific target group from Step 1 the toxic posts is directed toward. Select the single best match from A–H.
Step 4: Final Output (Q4)
Produce the final output as:
- If Q2 = 0 → no toxicity → output: `0;`
- If Q2 = 1, and Q3 has a clear target → output: `1;[A–H]`
% \end{minted}
\end{Verbatim}
\end{tcolorbox}

\begin{tcolorbox}[title=Prompt 2: Concise Definition with English Examples, colback=white!97!gray, boxrule=0.5pt, breakable ]
\begin{Verbatim}[fontsize=\scriptsize,breaklines=true,breaksymbolleft={},breaksymbolright={}]
As an expert annotator, assess whether a comment is toxic and identify which target group (if any) it refers to. Define toxicity as ``a rude, disrespectful, or unreasonable comment that is likely to make you leave a discussion.''
Annotation Guide
1. Toxicity Decision (0 or 1):  
- `0`: No  - `1`: Yes 
2. Target Category (A–H):  
   Identify the main target group :
   - `A`: players - `B`: NPCs - `C`: game developer
   - `D`: platform moderators  - `E`: policymakers
   - `F`: identity groups - `G`: other game-related entities - `H`: unknown
Output Format
Return a single line JSON object with the following keys:{"toxicity": 0 or 1, "target": "A–H"}
Examples
1. input: ``This update offers too little content for its price.''  output: {"toxicity": 0, "target": "C"}
2. input: ``Otome fans are clueless idiots who turn every thread into a fight.''  output: {"toxicity": 1, "target": "A"}
3. input: ``That love interest is a disgusting loser, and I'm sick of seeing him.''  output: {"toxicity": 1, "target": "B"}
4. input: ``Can a mod explain why my post got removed?''  output: {"toxicity": 0, "target": "D"}
5. input: ``Women are too stupid to understand game design and should stay quiet.''  output: {"toxicity": 1, "target": "F"}
% \end{minted}
\end{Verbatim}
\end{tcolorbox}

\begin{tcolorbox}[title=Prompt 3: Concise Reasoning with English Examples, colback=white!97!gray, boxrule=0.5pt, breakable ]
\begin{Verbatim}[fontsize=\scriptsize,breaklines=true,breaksymbolleft={},breaksymbolright={}]
As an expert annotator, assess whether a comment is toxic and identify which target group (if any) it refers to.
Define toxicity as ``a rude, disrespectful, or unreasonable comment that is likely to make you leave a discussion.''
Annotation Guide
1. Toxicity Decision (0 or 1):
   - `0`: No  - `1`: Yes 
2. Target Category (A–H):  
   Identify the main target group :
   - `A`: players - `B`: NPCs - `C`: game developer
   - `D`: platform moderators  - `E`: policymakers
   - `F`: identity groups - `G`: other game-related entities 
   - `H`: unknown
Output Format
Return a single line JSON object with the following keys:
{"toxicity_reason": "[Write your reason here, less than 20 words.]", "toxicity": 0 or 1, "target_reason": "[Write your reason here, less than 20 words.]", "target": "A–H"}
Examples
1. input: ``This update offers too little content for its price.''
output: {"toxicity_reason": "Criticizes content and price without insulting anyone", "toxicity": 0, "target_reason": "The complaint concerns the game update and its monetization", "target": "C"}
2. input: ``Otome fans are clueless idiots who turn every thread into a fight.''
output: {"toxicity_reason": "Predicates a demeaning term of otome players", "toxicity": 1, "target_reason": "The abusive language targets the player community", "target": "A"}
3. input: ``That love interest is a disgusting loser, and I'm sick of seeing him.''
output: {"toxicity_reason": "Predicates a demeaning term of a fictional character rather than evaluating the writing", "toxicity": 1, "target_reason": "The insult is directed at an in-game character", "target": "B"}
4. input: ``Can a mod explain why my post got removed?''
output: {"toxicity_reason": "Requests an explanation without attacking moderators", "toxicity": 0, "target_reason": "The post concerns a moderation decision", "target": "D"}
5. input: ``Women are too stupid to understand game design and should stay quiet.''
output: {"toxicity_reason": "Uses a sexist insult against women", "toxicity": 1, "target_reason": "The attack targets women as an identity group", "target": "F"}
% \end{minted}
\end{Verbatim}
\end{tcolorbox}

\section{Error Analysis}
\label{appendix:error}

\begin{CJK}{UTF8}{gbsn}
\begin{table*}[t!]
\centering
\caption{Representative false-positive (FP) and false-negative (FN) error modes for the selected classifiers on Weibo and Reddit. Examples are rewritten by the authors.}
\label{table:merged_error_analysis}
\renewcommand{\arraystretch}{1.4}
\begin{threeparttable}
\scalebox{0.8}{ 
\begin{tabular}{@{}l l p{6.5cm} p{7.5cm}@{}}
\toprule
\textbf{Platform} & \textbf{Error Type} & \textbf{Common Trigger} & \textbf{Illustrative Example} \\
\midrule
\multirow{3}{*}{\textbf{Weibo}} 
& \textbf{FP} & Sarcasm and playful insults that appear hostile without directly attacking a target 
& ``\textbf{狗叠} 这波活动又要我掏钱，谢谢你啊'' \newline
\textit{Trans: ``Dog-Paper Games wants my money again this time. Thanks a lot.''} \\
\cmidrule(lr){2-4}
& \multirow{2}{*}{\textbf{FN}} 
& Homophone obfuscation and abbreviated profanity 
& ``\textbf{tmd}这群人真是没救了，天天在超话带节奏'' \newline
\textit{Trans: ``Damn it (\textbf{tmd}), these people are hopeless, stirring up drama in the Super Topic every day.''} \\
& & Background-culture allusion and insider derogatory labels 
& ``随便发个截图就被说是\textbf{耀祖}老婆'' \newline
\textit{Trans: ``Posted one screenshot and got called Manchild's wife.''} \\
\hline
\multirow{3}{*}{\textbf{Reddit}} 
& \multirow{2}{*}{\textbf{FP}} 
& Profanity used as emphasis rather than abuse 
& \texttt{``Okay but what the fuck was that ending''} \\
& & Sarcasm without a clearly abusive target 
& \texttt{``So glad they brought him back. Not.''} \\
\cmidrule(lr){2-4}
& \textbf{FN} 
& Genre-specific derogatory slang and sexualized insults 
& \texttt{``There are a few manwhores in this route.''} \\ 
\bottomrule
\end{tabular}
}
\end{threeparttable}
\end{table*}
\end{CJK}

We acknowledge that none of the models is perfect; therefore, we provide a detailed error analysis to understand their limitations and potential biases.

\mypara{Binary Toxicity Detection}
Our error analysis for the binary toxicity detection task shows that the selected classifiers still face recall-side challenges, especially for culturally specific or obfuscated toxic expressions, as illustrated in~\autoref{table:merged_error_analysis}.
Consistent with the detector evaluation in~\autoref{table:toxicity_binary}, the selected classifier achieves a recall, equivalently TPR, of $0.80$ on Weibo and $0.78$ on Reddit.
This corresponds to false-negative rates of $20.0\%$ and $22.0\%$, respectively.
The remaining false negatives are not random.
On Weibo, they are often caused by homophone obfuscation, abbreviated profanity, and background-culture allusions.
On Reddit, they more often involve genre-specific slang, sarcasm, and profanity whose abusive meaning depends on otome-specific context.

False positives also differ across platforms.
On Weibo, the model sometimes over-interprets sarcastic complaints about game companies as direct abuse.
On Reddit, profanity used for emphasis or humor can be incorrectly classified as toxicity.
These errors suggest that the main difficulty is not only toxicity detection in the abstract, but the lexical-cultural mismatch between general-purpose language understanding and community-specific otome discourse.

Although the classifiers have limitations, these limitations do not obscure the large platform gap observed in the full corpus: the toxicity rate is $22.20\%$ in the Weibo general otome community and $3.71\%$ in the Reddit general otome community.

\begin{table*}[t!]
\caption{Comparison of target misclassification errors with platform-specific triggers. Examples are rewritten by the authors.}
\label{table:comparison_target_errors}
\begin{threeparttable}
\centering
\renewcommand{\arraystretch}{1.3} 
\scalebox{0.8}{
\begin{tabular}{@{}l c p{6.2cm} c p{6.2cm}@{}} 
\toprule
\multirow{2}{*}{\textbf{Confusion Pair}} & \multicolumn{2}{c}{\textbf{Weibo}} & \multicolumn{2}{c}{\textbf{Reddit}} \\
\cmidrule(lr){2-3} \cmidrule(lr){4-5}
& \textbf{Errors (N, \%)} & \textbf{Trigger Example} & \textbf{Errors (N, \%)} & \textbf{Trigger Example} \\
\midrule
A~$\leftrightarrow$~C & 11 (42\%) & \texttt{Dogdie is scamming us with this banner again.} & 2 (16\%) & \texttt{``whales''\tnote{a} or ``f2p''\tnote{b} players.} \\
\hline
H~$\rightarrow$~(Any) & 7 (27\%) & \texttt{FUCK OFF, all of you!!!} & 3 (25\%) & \texttt{Hate That!} \\
\hline
B~$\leftrightarrow$~C & 4 (15\%) & \texttt{They butchered my fave's card art in this patch.} & 4 (33\%) & \texttt{The devs ruined my favourite LI again.} \\
\hline
F~$\leftrightarrow$~C & 2 (7\%) & \texttt{Men are always this stingy.}& 2 (16\%) & \texttt{Of course they wrote him as another arrogant jerk.} \\
\hline
Other & 2 (7\%) & -- & 1 (8\%) & -- \\
\bottomrule
\end{tabular}
}
\begin{tablenotes}
    \item[a] \footnotesize ``whales'' are a minority of players who spend very large sums of money.
    \item[b] \footnotesize ``F2P'' (Free-to-Play) are players who do not spend money on the game.
\end{tablenotes}
\end{threeparttable}
\end{table*}

\mypara{Target Group Identification}
For the secondary task of target group identification, we manually examine the target-group misclassified instances in our error-analysis subset to identify recurring confusion patterns, summarized in~\autoref{table:comparison_target_errors}.
The most frequent error mode in this subset is confusion between players (\texttt{A}) and the game company or developers (\texttt{C}), especially in complaints related to gacha mechanics.
This \texttt{A}~$\leftrightarrow$~\texttt{C} confusion appears more frequently on Weibo than on Reddit, possibly reflecting stronger grievance fusion in its fan culture.
Another shared error mode involves ambiguous targets (\texttt{H} $\rightarrow$ Any), where vague outbursts or insider slang prevent specific target attribution.

\mypara{Toxicity Granularity Analysis}
To assess whether our toxicity definition is overly broad, we randomly sample 100 model-labeled toxic posts per platform and manually categorize them into three types: direct hostility (personal attacks, slurs, harassment), aggressive disagreement (strong negativity directed at groups, companies, or characters rather than named individuals), and benign/false positive.
On Weibo, 68\% of the sampled posts constitute direct hostility, suggesting that most model-labeled toxic posts involve direct harmful expression.
On Reddit, aggressive disagreement accounts for a larger share of the sampled posts (46\%), which may be related to the framing of criticism as collective consumer grievances.
Only 2\% of the sampled Weibo posts and 4\% of the sampled Reddit posts are benign false positives, suggesting that the model-labeled toxic subset is largely composed of genuinely negative or hostile content.

\mypara{Thread-Level Validation}
To examine whether isolated-post annotation introduces systematic bias, we additionally sample 200 discussion threads and re-annotate the focal posts with full conversational context.
The contextual annotations show high agreement with the original isolated-post labels, with Cohen's $\kappa=0.91$ for binary toxicity and $\kappa=0.87$ for target groups.
The few disagreements mainly involve counter-speech, sarcasm, and ambiguous references to players or developers.
A qualitative review further shows that these threads predominantly center on debates over character or narrative choices (62.3\%), counter-hate responses (28.1\%), and grievance articulation (9.6\%).
These results suggest that isolated-post annotation is generally stable for our measurement goals, while we acknowledge that thread-level context can still enrich qualitative interpretation.

\section{Prompt Optimization and Ablation Studies}
\label{appendix:prompt_optimization}

\begin{figure}[t!]
    \centering
    \includegraphics[width=0.9\linewidth]{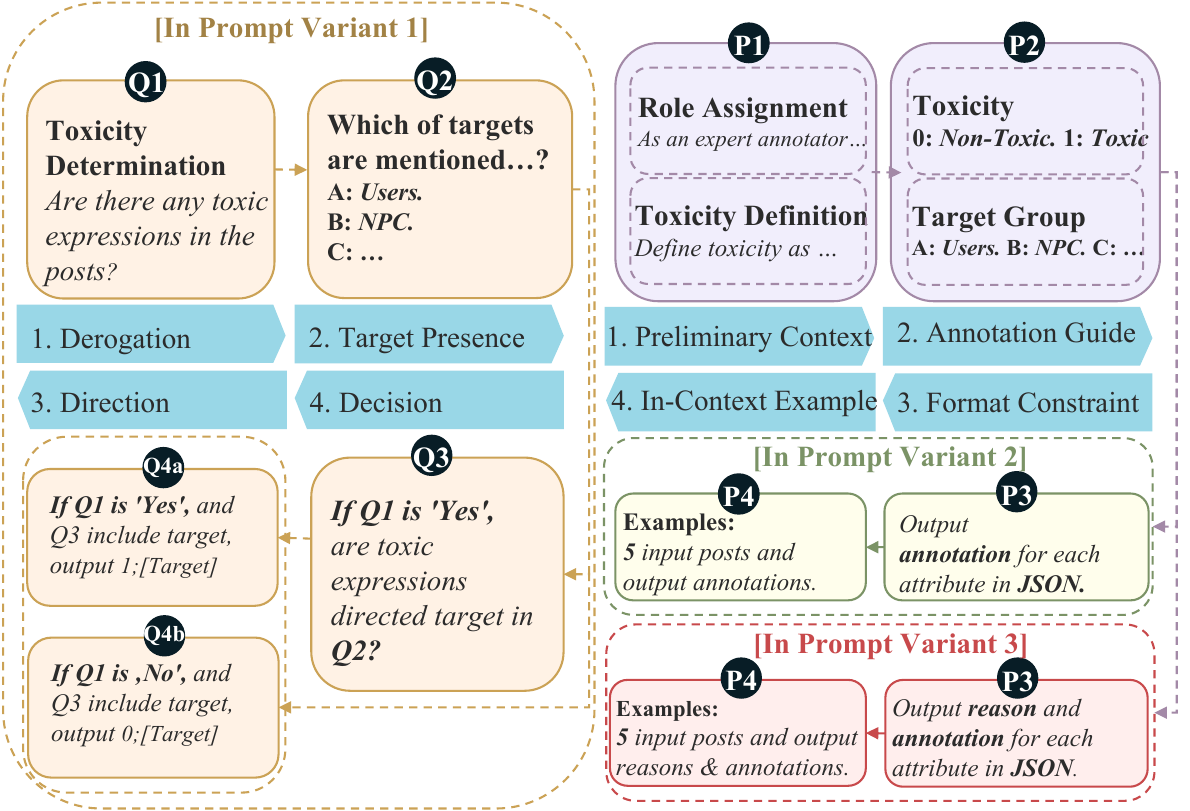}
    \caption{Logical flows for the CoT prompt and the Definition/Reasoning prompts.}
    \label{figure:prompt}
\end{figure}

\mypara{Prompt Templates}
We evaluate three prompt templates to classify toxicity, illustrated in~\autoref{figure:prompt}.
\begin{enumerate}
    \item \textbf{Chain-of-Thought (CoT):} We adapt a multi-step prompt from HateGuard~\cite{VGROCZH24} that guides the model through a sequential reasoning process.
    \item \textbf{Definition:} We implement a minimal prompt inspired by previous work~\cite{SWQBZZ25} that provides definitions and requires a direct JSON output without rationale.
    \item \textbf{Reasoning:} We extend the Definition prompt by requiring the model to output its reasoning for the classification, a technique shown to improve LLM stability~\cite{LAEZR24}.
\end{enumerate}

\mypara{Prompt Template Performance}
As shown in~\autoref{table:prompt_compare}, the Reasoning prompt (Prompt 3) outperforms the other prompt templates under the same evaluation setting, achieving an F1-score of 0.82 on Weibo and 0.78 on Reddit.

\mypara{Number of In-Context Examples}
Using the best-performing Reasoning prompt, we investigate the impact of the number of in-context examples ($E$). 
\autoref{table:nshot} shows that performance gains saturate around 5 to 15 examples.
Although 10-shot and 15-shot prompting yield marginally higher Reddit F1-scores, the improvement is small compared with the increased API cost.
Considering the trade-off between performance and the $\sim4\times$ increase in API cost from $E=0$ to $E=15$, we use 5-shot prompting as the default configuration.

\begin{table}[t!]
\centering
\caption{Impact of the number of examples ($E$).}
\label{table:nshot}
\tabcolsep 2pt
\scalebox{0.78}{
\begin{tabular}{@{}c|cccc|c|cccc|c@{}}
\toprule
\multirow{3}{*}{\textbf{\# Examples}} &
\multicolumn{5}{c}{\textbf{Weibo}} &
\multicolumn{5}{|c}{\textbf{Reddit}}  \\
\cmidrule(lr){2-11}
 &\multicolumn{4}{c|}{\textbf{Toxicity}} & \textbf{Target}  &
\multicolumn{4}{c|}{\textbf{Toxicity}} & \textbf{Target} \\
\cmidrule(lr){2-11}
& ACC & Prec. & Recall & F1 & ACC & ACC & Prec. & Recall & F1 & ACC \\
\midrule
0  &0.95 & 0.79 & 0.62 & 0.69 & 0.77 & 0.97 & 0.76 & 0.49 & 0.59 & 0.62\\
1  & 0.96 & 0.80 & 0.65 & 0.72 & 0.84 & 0.97 & 0.74 & 0.52 & 0.61 & 0.70\\
3  & 0.96 & 0.83 & 0.75 & 0.79 & 0.84 & 0.98 & 0.77 & 0.63 & 0.69 & 0.81\\
5  & 0.96 & \textbf{0.84} & \textbf{0.80} & \textbf{0.82} & 0.89 & 0.98 & 0.78 & \textbf{0.78} & 0.78 & \textbf{0.85}\\
10& \textbf{0.97} & 0.83 & 0.78 & 0.80 & 0.89 & 0.98 & 0.79 & 0.77 & \textbf{0.79} & \textbf{0.85}\\
15& \textbf{0.97} & \textbf{0.84} & \textbf{0.80} & \textbf{0.82} & \textbf{0.90} & 0.98 & \textbf{0.80} & \textbf{0.78} & \textbf{0.79} & \textbf{0.85}\\
\bottomrule
\end{tabular}
}
\end{table}

\begin{table}[t!]
\centering
\caption{Toxicity detection performance under three prompt templates.}
\label{table:prompt_compare}
\tabcolsep 2pt
\scalebox{0.74}{
\begin{tabular}{@{}l|cccc|c|cccc|c@{}}
\toprule
\multirow{3}{*}{\textbf{Prompt}} &
\multicolumn{5}{c}{\textbf{Weibo }} &
\multicolumn{5}{|c}{\textbf{Reddit }}  \\
\cmidrule(lr){2-11}
 &\multicolumn{4}{c|}{\textbf{Toxicity}} & \textbf{Target}  &
\multicolumn{4}{c|}{\textbf{Toxicity}} & \textbf{Target} \\
\cmidrule(lr){2-11}
& ACC & Prec. & Recall & F1 & ACC & ACC & Prec. & Recall & F1 & ACC \\
\midrule
1. CoT & 0.95 & 0.80 & 0.64 & 0.71 & 0.74 & 0.97 & 0.73 & 0.52 & 0.60 & 0.44  \\
2. Definition & 0.95 & 0.79 & 0.62 & 0.69 & 0.77 & 0.97 & 0.76 & 0.49 & 0.59 & 0.62 \\
3. Reasoning & \textbf{0.96} & \textbf{0.84} & \textbf{0.80} & \textbf{0.82} & \textbf{0.89} & 
           0.99 & \textbf{0.78} & \textbf{0.78} & \textbf{0.78} & \textbf{0.85}\\
\bottomrule
\end{tabular}
}
\end{table}

\section{Target Composition}
\label{appendix:event5_evolution}

In August 2024, several rappers publicly disparage otome players on Weibo, which coincides with a +14.89\% surge in toxicity.
\autoref{table:event5_evolution} tracks the day-by-day shift in target composition. 
Identity-group targeting (F) dominates in the first two days (38.1\%) but steadily declines as developer-targeted toxicity (C) rises sharply from 7.8\% to 19.5\%.
Player-on-player toxicity (A) remains stable at $\sim$27--30\% throughout, consistent with persistent internal friction.
This pattern shows qualitative target shifts rather than merely a quantitative spike.

\begin{table}[t!]
\centering
\caption{Target composition evolution during Event~\#5 (rapper external attack, August 2024) in the Weibo general otome community.}
\label{table:event5_evolution}
\scalebox{0.80}{
\renewcommand{\arraystretch}{1.15}
\begin{tabular}{@{}lcccc@{}}
\toprule
\textbf{Period} & \textbf{A (Players)} & \textbf{C (Developers)} & \textbf{F (Identity)} & \textbf{$n$} \\
\midrule
Day 1--2  & 26.6\% & 7.8\%  & 38.1\% & 451 \\
Day 3--4  & 26.8\% & 19.1\% & 35.9\% & 298 \\
Day 5--7  & 30.2\% & 19.5\% & 32.0\% & 169 \\
\bottomrule
\end{tabular}
}
\end{table}

\begin{table*}[t!]
\centering
\caption{Distribution of the four categories across target groups. 
Each cell shows the number of potential-coordination clusters, with row-wise percentages in parentheses.}
\label{table:attack_strategy_crosstab}
\scalebox{0.75}{
\begin{tabular}{@{}l rccccccc@{}}
\toprule
 & & \multicolumn{7}{c}{\textbf{Distribution of Target Groups: Count (\% of Row Total)}} \\
\cmidrule(l){3-9}
\textbf{Cluster Category} & \textbf{\# Clusters (\%)} & \textbf{Developers (C)} & \textbf{Unknown (H)} & \textbf{Players (A)} & \textbf{NPCs (B)} & \textbf{Identity Grp (F)} & \textbf{Moderators (D)} & \textbf{Other (G)} \\
\midrule
\textit{(low scale,\ \ \ \ short persistence)} & 82 (42.93\%) & 65 (79.27\%) & 4 (4.88\%) & 9 (10.98\%) & 1 (1.22\%) & 2 (2.44\%) & 0 (0.00\%) & \textbf{1 (1.22\%)} \\
\textit{(high scale, \ short persistence)} & 13 (6.81\%) & \textbf{12 (92.31\%)} & 0 (0.00\%) & 0 (0.00\%) & 0 (0.00\%) & \textbf{1 (7.69\%)} & 0 (0.00\%) & 0 (0.00\%) \\
\textit{(low scale,\ \ \ long persistence)} & 62 (32.46\%) & 30 (48.39\%) & 12 (19.35\%) & \textbf{11 (17.74\%)} & 4 (6.45\%) & 0 (0.00\%) & \textbf{5 (8.06\%)} & 0 (0.00\%) \\
\textit{(high scale, long persistence)} & 34 (17.80\%) & 16 (47.06\%) & \textbf{11 (32.35\%)} & 3 (8.82\%) & \textbf{3 (8.82\%)} & 0 (0.00\%) & 1 (2.94\%) & 0 (0.00\%) \\
\midrule
\textbf{Total} & \textbf{191 (100.00\%)} & \textbf{123 (64.40\%)} & \textbf{27 (14.14\%)} & \textbf{23 (12.04\%)} & \textbf{8 (4.19\%)} & \textbf{3 (1.57\%)} & \textbf{6 (3.14\%)} & \textbf{1 (0.52\%)} \\
\bottomrule
\end{tabular}%
}
\end{table*}

\begin{figure}[t!]
    \centering
    \includegraphics[width=1\linewidth]{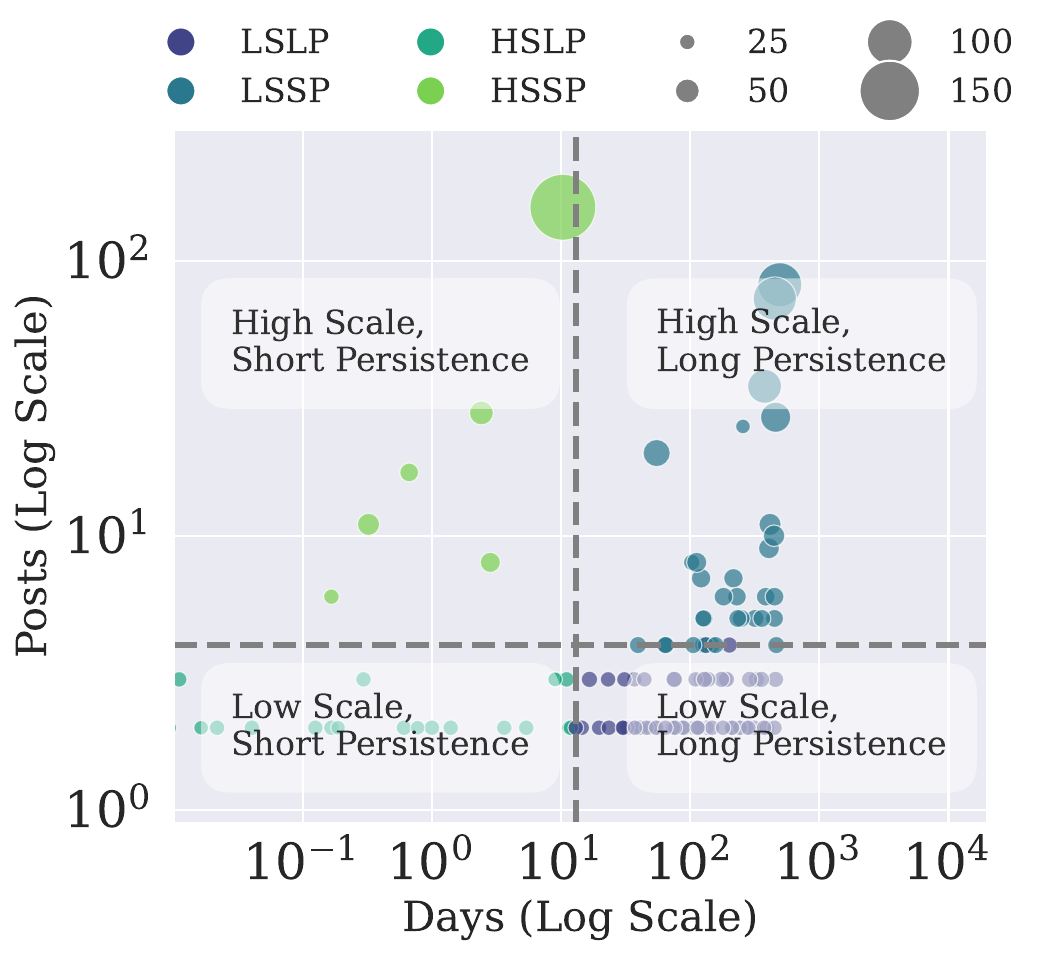}
\caption{Distribution of potential-coordination clusters by scale (posts) and persistence (days), with point size denoting the number of contributing accounts.}
    \label{figure:attack_taxonomy}
\end{figure}

\section{RQ1 Validation Analyses}
\label{appendix:rq1_validation}

\subsection{Interaction Regression Details}
\label{appendix:interaction_regression}

\begin{table}[t!]
\centering
\caption{Regression results of interaction features on toxicity across communities.
\colorbox{poscoef}{\textcolor{black}{Green}} = positive correlation (promotes toxicity), \colorbox{negcoef}{\textcolor{black}{Red}} = negative correlation (suppresses toxicity).
Significance: \textbf{*} $p<.05$, \textbf{**} $p<.01$, \textbf{***} $p<.001$.}
\label{table:regression_interactions}
\scalebox{0.8}{
\renewcommand{\arraystretch}{1.25}
\begin{tabular}{@{}clcc@{}}
\toprule
\textbf{Platform} & \textbf{Community} & \textbf{Like Effect} & \textbf{Comment Effect} \\
\midrule
\multirow{6}{*}{\textbf{Weibo}} 
& General Game Community & \cellcolor{poscoef}0.0165 (***) & \cellcolor{negcoef}-1.71 (***) \\
& General Otome Community & \cellcolor{poscoef}0.0275 (***) & \cellcolor{negcoef}-1.79 (***) \\
\cmidrule(lr){2-4}
& \textit{Genshin Impact} & \cellcolor{negcoef}-0.0293 (**) & \cellcolor{negcoef}-0.0382 (***) \\
& \textit{Love and Deepspace} & \cellcolor{negcoef}-0.2160 (***) & \cellcolor{negcoef}-2.01 (***) \\
& \textit{Mr. Love} & \cellcolor{negcoef}-0.0279 (***) & \cellcolor{negcoef}-0.4680 (***) \\
& \textit{Tears of Themis} & \cellcolor{negcoef}-0.5090 (***) & \cellcolor{negcoef}-0.3280 (***) \\
\midrule
\multirow{6}{*}{\textbf{Reddit}} 
& General Game Community & \cellcolor{negcoef}-0.0148 (***) & \cellcolor{poscoef}0.0098 (***) \\
& General Otome Community & \cellcolor{poscoef}0.0331 (***) & \cellcolor{poscoef}0.1860 (*) \\
\cmidrule(lr){2-4}
& \textit{Genshin Impact} & \cellcolor{negcoef}-0.0056 (***) & \cellcolor{poscoef}0.0805 (***) \\
& \textit{Love and Deepspace} & \cellcolor{poscoef}0.0261 (**) & \cellcolor{poscoef}0.0757 (***) \\
& \textit{Mr. Love} & \cellcolor{negcoef}-0.1610 (*) & \cellcolor{poscoef}0.7840 (***) \\
& \textit{Tears of Themis} & \cellcolor{poscoef}0.2850 (**) & \cellcolor{poscoef}0.1610 (***) \\
\bottomrule
\end{tabular}
}
\end{table}

\mypara{Robustness Check}
Because player-targeted toxicity~(A) and developer-targeted toxicity~(C) can be difficult to distinguish in complaints about gacha mechanics, game updates, and community disputes, we verify that our cross-platform conclusions are robust to potential A$\leftrightarrow$C confusion through two analyses.
First, we perform bootstrap resampling (10{,}000 iterations) on the annotated ground truth to compute 95\% confidence intervals for target group proportions.
As shown in~\autoref{table:bootstrap_ci}, the confidence intervals for Players~(A) and Developers~(C) are non-overlapping on both platforms (Weibo: A~$[43.8, 50.6]\%$ vs.\ C~$[26.5, 33.2]\%$; Reddit: A~$[6.1, 13.3]\%$ vs.\ C~$[17.6, 27.9]\%$), suggesting that the main target-ranking pattern is not driven by sampling uncertainty in the annotated toxic posts.
Second, as a confusion-invariant robustness check, we merge A~(players) and C~(developers) into a single \textit{human-target} category and compare against B~(NPCs).
Under this aggregation, Weibo otome toxicity is overwhelmingly human-targeted (A+C\,=\,92.9\% among \{A,\,B,\,C\}), whereas Reddit otome toxicity is predominantly NPC-targeted (B\,=\,51.1\% among \{A,\,B,\,C\}).
This cross-platform divergence, real individuals or organizations vs.\ fictional characters, is invariant to any A$\leftrightarrow$C swaps and reinforces our core finding.

\mypara{NPC vs.\ Developer Toxicity}
A natural question is whether toxicity targeting NPCs~(B) and developers~(C) reflects fundamentally different phenomena or merely different framings of the same grievance.
To investigate this question, we manually compare 50 randomly sampled posts from each category and additionally examine thread-level co-occurrence patterns in the thread-linked subset.
The sampled NPC-targeted posts are predominantly characterized by parasocial frustration, such as narrative disappointment and dissatisfaction with character design.
In contrast, developer-targeted posts more often reflect consumer grievances about game operation, monetization, or content quality.
In the thread-level analysis, NPC-targeted toxic threads rarely co-occur with developer-targeted toxicity in the same discussion chain, whereas developer-targeted threads more often co-occur with player-targeted toxicity.
These results suggest that NPC-targeted toxicity is not merely a proxy for developer criticism, but often reflects a distinct character-centered form of toxicity.

\mypara{Broader-Community Sanity Check}
To examine whether the elevated toxicity observed in otome communities is merely a byproduct of gender composition or gaming discourse in general, we conduct a small sanity check on broader non-otome communities.
Specifically, we sample 100 posts from K-pop communities, representing a female-dominated entertainment fandom, and 100 posts from \textit{League of Legends} communities, representing a mixed-gender gaming community, on both Weibo and Reddit.
We then apply the same annotation protocol used in our main study.

As shown in~\autoref{table:broader_community_sanity}, these broader communities show lower toxicity in this small manually checked sample.
Only one toxic post is observed in each sampled platform-community pair.
We use this analysis only as a sanity check rather than as a population-level estimate.
The result suggests that the elevated toxicity observed in Weibo otome communities is unlikely to be explained solely by female-dominated fandom composition or gaming discourse in general, although larger cross-community sampling would be needed for a definitive comparison.

\begin{table}[t!]
\centering
\caption{Broader-community sanity check on non-otome communities.
We sample 100 posts from each comparison community on each platform and apply the same annotation protocol.}
\label{table:broader_community_sanity}
\scalebox{0.8}{
\renewcommand{\arraystretch}{1.15}
\begin{tabular}{@{}llcc@{}}
\toprule
\textbf{Platform} & \textbf{Community} & \textbf{Type} & \textbf{\# Toxic / \# Sampled} \\
\midrule
Weibo & K-pop & Female-dominated & 1 / 100 \\
Reddit & K-pop & Female-dominated & 1 / 100 \\
Weibo & \textit{League of Legends} & Mixed-gender gaming & 1 / 100 \\
Reddit & \textit{League of Legends} & Mixed-gender gaming & 1 / 100 \\
\bottomrule
\end{tabular}
}
\end{table}

\mypara{Bootstrap Confidence Intervals}
To verify the robustness of our target group analysis against sampling uncertainty in the annotated toxic posts, we perform bootstrap resampling (10{,}000 iterations) on the annotated ground truth dataset.
\autoref{table:bootstrap_ci} reports the point estimates and 95\% confidence intervals for each target group's proportion among toxic posts.

\begin{table}[t!]
\centering
\caption{Bootstrap 95\% confidence intervals (10{,}000 iterations) for target group proportions in annotated toxic posts.}
\label{table:bootstrap_ci}
\scalebox{0.75}{
\renewcommand{\arraystretch}{1.15}
\begin{tabular}{@{}lcc@{}}
\toprule
\textbf{Target} & \textbf{Weibo $\hat{p}$ [95\% CI]} & \textbf{Reddit $\hat{p}$ [95\% CI]} \\
\midrule
A (Players)      & 47.2\% [43.8, 50.6] & 9.7\%  [6.1, 13.3]  \\
B (NPCs)         &  6.0\% [4.2, 7.8]   & 33.9\% [27.3, 40.5] \\
C (Developers)   & 29.8\% [26.5, 33.2] & 22.8\% [17.6, 27.9] \\
D (Moderators)   &  1.9\% [0.9, 2.8]   &  2.3\% [0.4, 4.3]   \\
E (Policymakers) &  0.2\% [0.0, 0.6]   &  0.6\% [0.0, 1.6]   \\
F (Identity)     &  7.8\% [5.9, 9.6]   & 20.0\% [14.8, 25.2] \\
G (Other)        &  3.9\% [2.6, 5.2]   &  0.6\% [0.0, 1.6]   \\
H (Unknown)      &  2.8\% [1.7, 3.8]   & 12.7\% [8.3, 17.1]  \\
\bottomrule
\end{tabular}
}
\end{table}

\section{Sensitivity Analysis}
\label{appendix:campaign_sensitivity}

To assess the sensitivity of our potential-coordination
detection results to the cosine-similarity threshold, we conduct a robustness analysis around $\theta=0.80$. 
The main procedure flags 191 high-similarity toxic clusters, which we treat as potential-coordination
candidates.
For each cluster, we assign a cluster-level target group by aggregating the post-level target labels within the cluster.
Specifically, we consider only posts classified as toxic and assign the cluster to the target group that appears most frequently among those toxic posts.
When two target groups are tied or the target otherwise remains ambiguous, we assign the cluster to the unknown category~(H).
This rule ensures that each cluster contributes to exactly one target group in~\autoref{table:attack_strategy_crosstab}.
Under this aggregation rule, 64.40\% of the 191 candidate clusters target game developers, making developers the dominant target group.

We vary the cosine-similarity threshold around the main setting and manually inspect the resulting clusters.
This analysis shows that the qualitative pattern remains stable:
clusters targeting game developers remain the dominant category, and recurring accounts continue to appear across multiple high-similarity toxic clusters.
Lowering the threshold includes more loosely related posts around the same controversy, while raising the threshold retains more near-duplicate posts with greater textual similarity.
In our manual inspection, clusters detected at $\theta=0.80$ frequently contain near-identical wording, repeated target references, and temporally concentrated posting patterns, providing stronger observable signals consistent with potential coordination rather than isolated toxic comments.
Although the number of detected clusters varies across thresholds, game developers remain their primary target, and a small set of accounts repeatedly participates across multiple clusters.

We therefore use $\theta=0.80$ in the main analysis because it balances two competing goals.
A lower threshold increases recall but risks grouping independent reactions to the same event into a single cluster.
A higher threshold increases precision but misses potentially coordinated posts that use minor lexical variation to avoid appearing identical.

\section{Toxicity Detection Model Details}
\label{appendix:models_details}

In this section, we provide detailed descriptions of the models used in our study.

\mypara{General-Purpose Detectors}
\begin{itemize}
    \item \textbf{Perspective API:} A widely used API fine-tuned on community forum data, offering multilingual support through checkpoints released in 2023~\cite{LTTSGMV22}.
    
    \item \textbf{OpenAI Moderation API:} A commercial API that uses a model distilled from GPT-4o for policy alignment, supporting nearly 40 languages~\cite{OpenAIModerationTooling, O24}.
    
    \item \textbf{COLD:} A RoBERTa-based classifier trained specifically on the Chinese Offensive Language Dataset, representing a strong baseline~\cite{DZSZMMH22}.
\end{itemize}

\mypara{LLM-Driven Detectors}
\begin{itemize}
    \item \textbf{DeepSeek-V3:} A Mixture-of-Experts large language model from DeepSeek, noted for its strong performance on Chinese language benchmarks~\cite{deepseekv3}.
    
    \item \textbf{DeepSeek-R1-Distill-Qwen-14B:} A distilled version of the DeepSeek-R1 series, built upon the Qwen2.5-14B architecture.
This model is designed to deliver efficient performance for reasoning, math, and code tasks~\cite{DeepSeekR1DistillQwen14B}.
    
    \item \textbf{GPT-4o:} OpenAI's flagship multimodal model.
It is natively designed to process a combination of text, audio, and visual inputs~\cite{GPT4o}.
    
    \item \textbf{GPT-4o mini:} The smaller, more efficient, and cost-effective counterpart to GPT-4o.
It is optimized for tasks requiring high throughput and lower latency~\cite{GPT-4o-mini}.
\end{itemize}

\section{Evasion-Aware Detection}
\label{appendix:gov:evasion}

\autoref{section:Linguistic-Strategies} shows that nearly 40\% of toxic spans on Weibo employ variation strategies, and that these evasion-laden posts attract significantly more engagement than direct toxic posts (mean 26.76 vs.\ 8.37 comments).
This means current moderation disproportionately misses the most visible toxic content.
Our taxonomy (\autoref{table:linguistic_strategies_taxonomy}) suggests four corresponding pre-processing modules that can be layered before existing classifiers.

The first targets community-specific slang and memes (26.49\% of Weibo toxic spans).
From our corpus we extract approximately 320 Chinese and 85 English toxic terms absent from standard lexicons.
This seed lexicon can be kept current through a lightweight weekly pipeline: extract high-frequency novel tokens from Super Topics or subreddits, query an LLM for contextual toxicity assessment, and surface candidates for human review.

The second addresses Pinyin and letter-code abbreviations (6.58\% on Weibo), such as ``sb'' for ``shabi'' (idiot) or ``tmd'' for ``ta ma de.'' A mapping table from common initial sequences to their most probable vulgar expansions, weighted by corpus frequency, allows the system to expand abbreviations and re-score them---flagging posts where the expanded form triggers the classifier but the abbreviated form does not.
The same logic applies to English abbreviations like ``stfu.''

The third handles homophone and visual substitution (5.03\% on Weibo).
A phonetic canonicalization layer that converts text to Pinyin and detects collisions with known vulgar terms, combined with Unicode confusable mappings for glyph-level obfuscation, can normalize these substitutions before classification.

The fourth covers emoji substitution (1.79\% on Weibo), where emojis replace offensive words.
An emoji-in-context classifier, fine-tuned on our annotated toxic spans, can predict whether a given emoji functions as a toxic stand-in based on surrounding text.

Deployed together as a normalization layer, these modules target the specific evasion channels our data reveals and could substantially reduce the false negative rate (currently 19.8\% on Weibo).

\section{Early-Warning Indicators for Potential Coordination}
\label{appendix:gov:earlywarning}

The 191 potential-coordination clusters flagged in~\autoref{section:case_study_weibo} exhibit observable structural signals, including lexical homogeneity, temporal concentration, recurring low-history accounts, and concentration on a single target.
Based on these observations, we outline five candidate signals that may help moderators prioritize emerging patterns of potential coordination for human review.

The first signal is content homogeneity, measured using rolling pairwise TF-IDF cosine similarity within a sliding six-hour window.
Because our main analysis uses a similarity threshold of $0.80$, a lower threshold such as $0.60$ could be evaluated as a preliminary alert threshold.
The second signal is new-account influx, measured as the proportion of posts authored by accounts less than 30 days old.
Our account-participation analysis shows that several contributing accounts were created shortly before the relevant posting periods.
An influx exceeding twice the community baseline could therefore warrant human review.
The third signal is toxicity velocity, measured as the hourly change in toxicity ratio.
Event~\#5's 72-hour increase from the baseline to $+14.89\%$ illustrates the steep trajectories that may precede major toxicity surges.
The fourth signal is target concentration, measured by whether one target group accounts for more than 70\% of toxic posts in a window relative to the baseline distribution in~\autoref{figure:target}.
The fifth signal is high-reach-account activity, motivated by our observation that one verified account contributed to six potential-coordination clusters.
This observation does not establish that the account mobilized other participants.

These signals are intended to prioritize content for human review rather than establish coordination automatically.
As a possible tiered design, the co-activation of two signals could trigger expedited human review, while three or more could trigger temporary and reversible visibility reduction pending review.
The thresholds should be calibrated separately for each community using manually reviewed potential-coordination clusters as provisional positive examples and randomly sampled non-candidate periods as negative examples.
Such calibration should prioritize high precision while targeting recall of at least $0.80$ to limit false alarms.

This framework is tailored to the potential-coordination patterns observed in our data, which are characterized by lexical homogeneity, temporal concentration, and participation by low-activity accounts.
More subtle forms of potential coordination involving semantic diversity or slow posting patterns would require complementary detection approaches.

\section{Platform-Specific Intervention Experiments}
\label{appendix:gov:experiments}

The opposing toxicity--engagement dynamics on Weibo (suppression: toxic posts receive fewer likes and comments) and Reddit (amplification: toxic posts attract more comments) call for distinct intervention strategies.
We outline two concrete A/B experimental designs.

\mypara{Weibo}
We hypothesize that excluding posts flagged as toxic (score $>$0.70, using an enhanced classifier incorporating the evasion-aware modules from \refappendix{appendix:gov:evasion}) from the ``hot posts'' feed and Super Topic homepage, while keeping them accessible via direct search, will reduce community-level toxicity without suppressing non-toxic engagement.
The experiment can be run across matched Super Topic pairs of similar size and baseline toxicity over a minimum of 8 weeks (sufficient to capture at least one natural toxicity event cycle, per our temporal analysis in \autoref{section:temporal}).
Primary outcomes are weekly toxicity ratio, non-toxic engagement volume, and user retention.
Our finding that evasion-strategy posts receive 3.2$\times$ more engagement suggests that visibility is a key amplification mechanism, so de-amplification should yield a meaningful effect.
Importantly, this approach does not remove content but reduces algorithmic promotion, preserving users' ability to seek out specific discussions.

\mypara{Reddit}
Since 22.75\% of Reddit otome toxic posts target developers and toxic posts correlate positively with comment counts (\autoref{section:RQ1}), we hypothesize that introducing a weekly pinned ``Developer Feedback'' megathread with structured templates (issue description, expected behavior, suggested fix) will redirect destructive toxicity into constructive criticism.
In the treatment subreddit, automod directs posts containing developer-critical keywords to the megathread; the control retains the status quo.
Primary outcomes are the proportion of developer-targeted toxic posts (Target~C) in general threads and a constructiveness score assessed via LLM-based evaluation.

Both designs are intentionally lightweight and reversible, making them practical for community moderators to pilot without platform-level engineering changes.

\end{document}